\documentclass[a4paper,11pt]{article}
\usepackage{jheppub}

\newcommand{\GeV}{{\,\mathrm{GeV}}}
\newcommand{\rL}{{\mathrm{L}}}
\newcommand{\rR}{{\mathrm{R}}}
\newcommand{\rT}{{\mathrm{T}}}
\newcommand{\rd}{{\mathrm{d}}}
\newcommand{\ri}{{\mathrm{i}}}
\newcommand{\rs}{{\mathrm{s}}}
\newcommand{\TF}{T_\mathrm{F}}
\newcommand{\CA}{C_{\mathrm{A}}}
\newcommand{\CF}{C_{\mathrm{F}}}

\newcommand{\gs}{g_\rs}
\newcommand{\gL}{g_\rL}
\newcommand{\gY}{g_Y}
\newcommand{\gZ}{g_{Z^0}}

\title{Real effective potentials for phase transitions in models with extended scalar sectors}

\author[a]{K\'aroly Seller,}
\author[b]{Zsolt Sz{\'e}p}
\author[a,c]{and Zolt\'an Tr{\'o}cs\'anyi}

\affiliation[a]{Institute for Theoretical Physics, ELTE E{\"o}tv{\"o}s Lor\'and University,\\
P\'azm\'any P{\'e}ter s{\'e}t\'any 1/A, H-1117 Budapest, Hungary}
\affiliation[b]{ELKH-ELTE Theoretical Physics Research Group,\\
P\'azm\'any P{\'e}ter s{\'e}t\'any 1/A, H-1117 Budapest, Hungary}
\affiliation[c]{Institute of Experimental Physics, Faculty of Sciences and Technology, University of Debrecen,
Bem tér 18/A, H-4026 Debrecen, Hungary}

\emailAdd{karoly.seller@ttk.elte.hu}
\emailAdd{szepzs@achilles.elte.hu}
\emailAdd{zoltan.trocsanyi@cern.ch}

\abstract{ 
The effective potential obtained by loop expansion is usually not real in the range of field values explored by its minima during a phase transition. 
We apply the optimized perturbation theory in a fixed gauge to singlet scalar extensions of the Standard Model in order to calculate a one-loop effective potential that is real by construction. 
We test this computational scheme by comparing such a potential obtained in Landau gauge to that derived based on the Higgs pole mass. 
We carry out the latter construction by imposing physical renormalization conditions, which yields a potential without residual regularization scale dependence.
We use our effective potential to study the parameter dependence of the critical temperatures in a two-step phase transition of the form $(0,0)\to (0,w')\to (v,w)$ that occurs for decreasing temperature in scalar extensions of the SM with two vacuum expectation values $v$ and $w$.}

\begin{document}
\allowdisplaybreaks
\maketitle

\section{Introduction}

We know from measurements of anisotropies in the cosmic microwave background \cite{Jungman:1995bz,Planck:2018vyg}, Big Bang nucleosynthesis \cite{Fields:2019pfx}, and direct searches \cite{AMS:1999tha} that the Universe contains baryons and trace amounts of anti-baryons as ordinary matter.
From the value of the baryon-to-photon ratio $\eta \simeq 6\cdot 10^{-10}$ \cite{Workman:2022ynf} we know that in the early Universe there should have been approximately one extra baryon for a billion baryon and anti-baryon pairs.
These pairs could annihilate leaving the extra baryons to survive and form the Universe we know today.
This observation cannot be explained by the Standard Model of particle physics (SM), as it does not predict sufficiently large violation of the CP symmetry \cite{Jarlskog:1985ht,Gavela:1993ts} required by baryogenesis \cite{Sakharov:1967dj}.
Nevertheless, it is an intriguing question whether the known baryon asymmetry can be explained by  a suitable extension of the SM.

There are two main directions to explain generation of baryon asymetry in particle physics. 
One is the mechanism of baryogenesis \cite{Riotto:1998bt,Cline:2006ts} in which sphalerons provide the necessary baryon number violating processes. 
Baryogenesis requires a first order phase transition as part of Sakharov's conditions \cite{Sakharov:1967dj}, which is strong enough to prevent the washout of the created baryon number.
Another one is the mechanism of leptogenesis \cite{Davidson:2008bu,Bodeker:2020ghk}, which seeks to create a lepton number asymmetry by involving heavy neutral leptons (singlet states under the SM gauge group), then turns this lepton asymmetry into a baryon asymmetry by standard processes \cite{Harvey:1990qw}.
For leptogenesis one needs to know the temperature range where the relevant phase transition or transitions happened and follow the thermal evolution of the masses within this range.

Beyond the SM theories may provide a solution to the baryon asymmetry as they generally incorporate new degrees of freedom.
In particular, models with extended scalar sectors have interesting phenomenology regarding phase transitions in the early Universe \cite{Espinosa:2011ax,Curtin:2014jma,Katz:2014bha,Chiang:2017nmu}.
A first order phase transition requires a potential barrier between the symmetric and the symmetry breaking ground states that can either be generated already at tree level by a suitable Lagrangian \cite{Espinosa:2011ax}, or at loop level by the thermal contributions of bosonic degrees of freedom \cite{Carrington:1991hz}.
It is known that in the SM, the thermal effects do not produce a sufficiently strong first order electroweak phase transition (EWPT) as the physical Higgs boson is too heavy \cite{Csikor:1998eu, DOnofrio:2014rug}.
In extensions of the SM, additional bosons coupled to the Brout-Englert-Higgs (BEH) field can strengthen this transition \cite{Anderson:1991zb}.

The simplest scalar extension of the SM adds one gauge singlet scalar.
For a $\mathbb{Z}_2$ symmetric scalar a tree-level potential barrier is excluded between the electroweak minimum and an additional minimum with a non-vanishing singlet vacuum expectation value (VEV) \cite{Espinosa:2011ax}, while thermal effects are generally too weak to strongly affect the thermally generated symmetry breaking. 
In these extensions a strongly first order phase transition is only possible if the singlet VEV vanishes in the $T=0$ broken phase (see e.g.~Ref.~\cite{Curtin:2014jma,Ghorbani:2020xqv} for a real singlet and Ref.~\cite{Katz:2014bha} for a complex one).
Hence, baryogenesis is ruled out if the second VEV is also nonzero.
Nevertheless, leptogenesis is still possible in such models if heavy neutral leptons (HNLs) exist. 
For example, in the superweak extension of the SM (SWSM) \cite{Trocsanyi:2018bkm} the HNLs are the right-handed neutrinos, which receive their masses from the finite VEV of a singlet scalar through the BEH mechanism.
To study such scenarios of leptogenesis, we have to find the critical temperatures of the phase transition of the model.

In quantum field theory the effective potential is the relevant tool for thermodynamic calculations and the study of phase transitions. 
Evaluated at homogeneous field configuration from the Legendre transform of the generating functional, it is convex, infrared (IR) finite, and real\footnote{In field theory the imaginary effective potential is usually associated with the decay of the metastable vacuum \cite{Coleman:1985rnk}, which does not concern our study.}.
Some functional approaches used to compute the effective potential indeed satisfy these expectations \cite{Wipf:2021mns}.
However, for models with non-convex (double-well) classical potentials the effective potential obtained by loop expansion at a fixed order is not convex in general \cite{Jackiw:1974cv}.
Additionally, beyond leading order the effective potential is complex for small field values, and even worse, in the presence of massless modes it becomes infrared divergent starting at three-loops (and its derivatives already at lower loops \cite{Ford:1992mv, Martin:2014bca}). 
In this case the elimination of IR divergences from the effective potential can be achieved by resummation of the Goldstone self-energy \cite{Martin:2014bca,Elias-Miro:2014pca}. 
In the SM the calculation of the effective potential was carried out recently at two-loop \cite{Ekstedt:2020abj} and even at three-loop order \cite{Ekstedt:2020abj,Martin:2013gka, Martin:2017lqn, Espinosa:2017aew}.
Furthermore, the infrared instabilities can also be treated with dimensional reduction, where the heavy modes in the theory are integrated out and the problematic low energy theory is studied nonperturbatively on the lattice \cite{Kajantie:1995dw,Kajantie:1995kf}.
For recent applications of the method in scalar extensions of the SM see Refs.~\cite{Schicho:2021gca, Niemi:2021qvp}.

The origin of the imaginary part of the effective potential arising in between its minima within the conventional loop expansion is a consequence of the non-convexity of the classical potential \cite{Fujimoto:1982tc}.
In Ref.~\cite{Weinberg:1987vp} the authors interpreted the imaginary part of the one-loop result as the decay rate of a specific Gaussian state centered about field values where the classical potential was concave.
Indeed, there are no stable homogeneous equilibrium states in this region \cite{Cooper:1996ii}.
However, the loop expansion of the effective potential is reliable only when the defining path integral is dominated by a single saddle point, which is the case for convex classical potentials.
For models with a non-convex classical potential the usual expansion fails to correctly approximate the exact effective potential as competing saddle points are usually not taken into account \cite{Rivers:1983sq,Rivers:1987hi}.
Mathematically the issue can also be formulated as the non-interchangeability of the loop expansion and the analytic continuation of a parameter that controls the convexity, as illustrated in Sec.~13.5 of Ref.~\cite{Rivers:1987hi}.
A modified loop expansion, which takes into account two saddle points, was considered in Ref.~\cite{Fujimoto:1982tc}. 
This procedure leads to an effective potential that is convex and real.
Also, a generalized effective potential that is real and that can be computed perturbatively was constructed in Ref.~\cite{Cahill:1993mg} by coupling the external current to a polynomial of the field.

Motivated by the above considerations, our aim in this paper is to study the tem\-per\-a\-ture-driven phase transition of scalar extended models with $\mathbb{Z}_2$ symmetry as a first step in a  phenomenological investigation of leptogenesis. 
In particular, we plan to find the ranges of the critical temperatures in a two-step phase transition where the singlet acquires a non-zero VEV at high temperature, and remains finite even at temperatures below the EWPT (for vanishing singlet VEV, see e.g.~Ref.~\cite{Athron:2022jyi}).

In order to calculate the one-loop effective potential, which is enough to estimate the range of the critical temperature available in a given model, we employ the optimized perturbation theory (OPT) of Ref.~\cite{Chiku:1998kd}.
This computational scheme was applied mainly in scalar models where it was shown to preserve Ward identities (Goldstone theorem) and renormalizability even at higher orders in the loop expansion \cite{Chiku:2000eu} as it preserves the symmetries of the Lagrangian.
For an application of the method to describe the BEC-BCS crossover in a non-renormalizable fermionic model, see Ref.~\cite{Duarte:2017zdz}.
The OPT scheme yields an infrared finite and real effective potential in a sufficiently wide, relevant region of the scalar fields involved. 
The reality of the OPT potential is ensured by the convexity of the classical potential, which in turn ensures that the conventional loop expansion is reliable.
Thus the application of the OPT solves the problem of the conventional loop expansion which does not give a real effective potential at vanishing field value, even {\em at high temperature when it becomes the stable ground state of the system}.

The paper is organized as follows. 
In Sec.~\ref{sec:SM_pot} we deal with the SM one-loop effective potential at zero temperature.
Using Landau gauge, we construct an IR finite effective potential which is used as a benchmark for the effective potential calculated with the OPT method. 
We also discuss the physical parametrization of these potentials.
In Sec.~\ref{sec:sSM_pot} we discuss the tree-level potential and construct at zero temperature the one-loop OPT effective potential in a complex scalar extension of the SM. 
The parametrizations of the one-loop OPT potential of this and the SWSM model are discussed in Sec.~\ref{sec:param_sSM_pot}.
In Sec.~\ref{sec:finT_pot} we provide the one-loop thermal corrections to the OPT potential and investigate the phase transition of the SWSM model. 
We present several auxiliary computations in the Appendices: the Higgs self-energy needed for the benchmark potential in App.~\ref{app:SE} and steps of the construction of the benchmark potential in App.~\ref{app:PoleBasedPot}, while the derivation of the thermal masses is discussed at length in App.~\ref{app:ThermalMasses}.

\section{Standard model effective potential at one loop \label{sec:SM_pot}} 

The SM scalar sector includes an SU(2)$_\mathrm{L}$ doublet, the BEH field, that in the spontaneously broken phase is parametrized in terms of real scalar fields as:
\begin{equation}
    \label{eq:BEHfield}
    \phi(x)=\frac{1}{\sqrt{2}}
    \begin{pmatrix}
        \phi_3(x)+\ri\phi_4(x) \\
        v + h(x) + \ri\phi_2(x)
    \end{pmatrix}\,.
\end{equation}
Here $\phi_i(x)$ and $h(x)$ are the Goldstone and Higgs boson fields and $v$ is the constant background field about which the Higgs field oscillates.

The associated scalar potential
\begin{equation}
  \label{eq:V_SM_cl}
    \mathcal{V}(\phi)=\mu^2 |\phi|^2 + \lambda |\phi|^4\
\end{equation}
depends on the mass parameter $\mu^2$ and self-coupling $\lambda$. 
The classical potential is obtained upon neglecting all oscillating fields in the BEH doublet, and taking the potential to be a function of the homogeneous background field $v$:
\begin{equation}
\label{eq:V_SM_cl2}
    V_\mathrm{cl}(v;\mu^2)=\frac{\mu^2}{2}v^2 + \frac{\lambda}{4}v^4\,.
\end{equation}
Here we explicitly indicated the dependence on the mass squared parameter for later convenience.
In order for this potential to have a non-trivial minimum, we require that $\mu^2<0$ and $\lambda>0$. 
These tree-level parameters are determined by the VEV $v_0$ and the mass $M_h$ of the Higgs boson (c.f.~Table~\ref{tab:SMParameters}): $\mu^2=-M_h^2/2$ and $\lambda=M_h^2/(2v_0^2)$.
The positivity of $\lambda$ also implies that the resulting classical potential is bounded from below.
\begin{table}
    \centering
    \begin{tabular}{|l|c|c|c|c|c|}
        \hline
        Parameters & $v_0$ & $M_h$ & $M_t$ & $M_Z$ & $M_W$ \\ \hline
        Values & $246.22\GeV$ & $125.25\GeV$ & $172.69\GeV$ & $91.188\GeV$ & $80.377\GeV$ \\
        \hline
    \end{tabular}
    \caption{Collection of SM mass and VEV values taken from Ref.~\cite{Workman:2022ynf}. At leading order the VEV is obtained from the value of the Fermi constant as $v_0=(\sqrt{2}G_\mathrm{F})^{-1/2}$.}
    \label{tab:SMParameters}
\end{table}

The full effective potential is a sum of the classical potential and loop corrections, which we write formally as
\begin{equation}
    V^{[n]}_\mathrm{eff}(v;\mu^2)=V_\mathrm{cl}(v;\mu^2)+\sum_{\ell=1}^n V^{(\ell)}(v;\mu^2)\,.
    \label{eq:Veff}
\end{equation}
In particular, the first order correction is given by the sum of various one-loop contributions
\begin{equation}
 \label{eq:V_SM_1loop}
    V^{(1)}\big(v;\mu^2\big)=\sum_i n_i V_{\mathrm{CW}}\big(m_i^2(v;\mu^2)\big)\,,
\end{equation}
where each individual mode $i$ with multiplicity $n_i$ has the Coleman-Weinberg (CW) form
\begin{equation}
  \label{eq:V_SM_CW}
  V_{\mathrm{CW}}\big(m_i^2(v;\mu^2)\big)=\frac{s_im_i^4(v;\mu^2)}{64\pi^2}\left(\ln\frac{m_i^2(v;\mu^2)}{Q^2} - c_i+D_{\overline{\rm MS}}\right)\,.
\end{equation}
Here $m_i^2(v;\mu^2)$ is the tree-level background-dependent mass squared, and the overall sign is $s_i=-1$ for fermions and $s_i=1$ for bosons. 
The CW form \eqref{eq:V_SM_CW} is obtained using dimensional regularization (DR) and modified minimal subtraction scheme ($\overline{\text{MS}}$), with $Q$ as the regularization scale. 
The UV divergence proportional to $D_{\overline{\rm MS}} = -\varepsilon^{-1} + \gamma_{\rm E} - \ln(4\pi)$ is canceled by appropriate counterterms. 
The value of the constant $c_i$ appearing in \eqref{eq:V_SM_CW} is $3/2$ for fermions and scalars and $5/6$ for massive gauge bosons\footnote{The value is different as there are $d-1$ transverse modes and $(d-1)D_{\overline{\rm MS}}=3 D_{\overline{\rm MS}}+2+\mathcal{O}(\varepsilon)$.}.

In the SM all massive particles obtain their masses through the BEH mechanism.
In principle every particle with a mass contributes to the effective potential, but we can safely neglect all but the heaviest ones. 
Working in Landau gauge ($\xi=0$), in which the ghosts are massless, the background-dependent tree-level masses of the relevant particles are 
\begin{subequations}
\label{eq:SM_tree_m2}
  \begin{gather}
    \label{eq:SMmassesWZ}
    m_W^2(v) = \frac{\gL^2 v^2}{4}\,,\quad m_Z^2(v) = \frac{(\gL^2 + \gY^2)v^2}{4}\equiv \frac{\gZ^2 v^2}{4}\,, \\
    \label{eq:SMmassesTop}
    m_t^2(v) = \frac{y_t^2 v^2}{2}\,, \\
    \label{eq:SMmassesScalars}
    m_h^2(v;\mu^2) = \mu^2 + 3\lambda v^2\,,\quad m_{\phi_i}^2(v;\mu^2) \equiv m_G^2(v;\mu^2 )= \mu^2 + \lambda v^2\,,\quad\text{for }i=2,3,4\,,
\end{gather}
\end{subequations}
where $\gL$ and $\gY$ are the SU(2)$_\rL$ and U(1)$_Y$ couplings and $y_t$ is the Yukawa coupling of the top quark.
The multiplicities of these particles are $n_G=3n_h=3$, $n_W=2n_Z=6$ and $n_t=12$.

There are two issues regarding the effective potential that need to be addressed.
First, the second derivative of the one-loop effective potential, which is the curvature mass of the Higgs boson, is IR divergent in the minimum $v_0$ due to the massless Goldstone modes.
Secondly, as $\mu^2<0$, the Goldstone mass squared becomes negative for $v<\sqrt{-\mu^2/\lambda}\equiv v_0$ (c.f.~Eq.~\eqref{eq:SMmassesScalars}).
This makes the effective potential complex for field values $v<v_0$.
A complex potential would indicate instability, which we know is not the case, thus this feature is understood to be an unphysical artifact of the loop expansion of the effective potential.
We discuss these issues in detail in the following two sections.

\subsection{Effective potential with parametrization based on the Higgs pole mass}
\label{sec:PoleMassparametrization}

If one tries to construct the one-loop effective potential using renormalization conditions that preserve the tree-level values of the minimum and the curvature mass of the Higgs boson \cite{Anderson:1991zb}, one encounters a logarithmic IR divergence caused by the massless Goldstone modes, collectively denoted by $G$.
This is because, although the one-loop potential \eqref{eq:V_SM_CW} is finite for a vanishing mass, its second derivative is not: the bubble integral at vanishing external momentum $\mathcal{B}(p^2=0;m_G)$ is IR divergent for $m_G(v_0;\mu^2)=0$.
The standard way to avoid the appearance of this logarithmic divergence is to impose a renormalization condition on the pole mass $M_{\rm p}\equiv M_h$ (see e.g. \cite{Boyd:1992xn,Cline:1996mga,Delaunay:2007wb}) instead of the curvature mass of the Higgs boson. 
At the minimum $v_0$ of the effective potential the pole mass is given in terms of the self-energy $\Pi$ of the Higgs boson as \cite{Casas:1994us, Degrassi:2012ry}
\begin{equation}
  M_{\rm p}^2 = m_h^2(v_0;\mu^2) + \Pi(p^2=M^2_{\rm p};v_0)\,.
\end{equation}  
With this procedure, the bubble integral appearing in the curvature mass is replaced by $\mathcal{B}(p^2=M^2_{\rm p};m_G)$, which is finite for a vanishing mass because $p^2=M^2_{\rm p}$ acts as an IR regulator. 

In order to formulate the pole mass based parametrization of the one-loop effective potential, we add the following term to the tree-level Lagrangian of the Higgs field
\begin{subequations}
\label{eq:L_CT_notation}
\begin{equation}
\delta \mathcal{L}_{\rm SM}\supset \frac{\delta Z}{2}\partial_\mu h(x)\partial^\mu h(x) - \frac{\delta\mu^2}{2}\big(v+h(x)\big)^2 - \frac{\delta\lambda}{4}\big(v+h(x)\big)^4\,,
\end{equation}
which contains through the decomposition
\begin{equation}
\label{eq:cpl_decomp}
\delta Z = \delta_Z + \Delta Z, \quad \delta \mu^2 = \delta_{\mu^2} + \Delta\mu^2, \quad \delta \lambda = \delta_\lambda + \Delta\lambda,
\end{equation}
\end{subequations}
infinite counterterms and finite corrections to the tree-level couplings, both nonvanishing and scheme-dependent beyond tree level. Then, the usual relations between the bare couplings $\lambda_0$, $\mu_0^2$ and the renormalized ones are $\mu_0^2=\mu^2_{\rm R} + \delta_{\mu^2}$, $\lambda_0=\lambda_{\rm R}+\delta_\lambda$, with $\mu^2_{\rm R} = \mu^2 + \Delta \mu^2$ and $\lambda_{\rm R}=\lambda+\Delta\lambda$. Here $\mu^2$ and $\lambda$ are finite couplings that satisfy the tree-level relations given below \eqref{eq:V_SM_cl2} by construction and enter into the expression of the tree-level masses, while $\delta_Z$ and $\Delta Z$ represent the infinite and finite part of the wave function renormalization factor.

Working in a strict perturbative expansion, we choose the finite corrections to the tree-level parameters, $\Delta Z,\,\Delta {\mu^2}$, and $\Delta\lambda$, such that they remove all finite one-loop quantum corrections from the Higgs one- and two-point functions.
This treatment ensures that the pole mass, the residue at the pole, and the minimum of the potential do not change compared to their respective tree-level values.  
The counterterms $\delta_Z,\,\delta_{\mu^2}$ and $\delta_\lambda$, which remove UV divergences in the ${\overline{\rm MS}}$ scheme, are determined based on the Higgs self-energy given in App.~\ref{app:SE}. They cancel the divergences in the potential introduced in \eqref{eq:V_SM_1loop}.

The renormalized one-loop effective potential obtained with the above procedure, presented in details in  App.~\ref{app:PoleBasedPot}, is ($m_{h/G}(v)\equiv m_{h/G}(v;\mu^2)$)
\begin{eqnarray}
V^{[1]}(v)&=&\Omega + \frac{\lambda}{4}(v^2-v_0^2)^2 + \frac{1}{4}\sum_{i\ne G}
l_i\big(M_h^2/m_i^2(v_0)\big) (v^2-v_0^2)^2
\nonumber\\
&&+ \frac{n_G}{64\pi^2}m_G^4(v)\left(\ln\frac{m_G^2(v)}{m_h^2(v_0)} + \frac{3}{2}\right)
\nonumber\\
&&+\sum_{i\ne G} \frac{n_i s_i}{64\pi^2}\left[m_i^4(v)\left(\ln\frac{m_i^2(v)}{m_i^2(v_0)}-\frac{3}{2}\right) + 2 m_i^2(v_0)m_i^2(v)\right],
\label{eq:V_SM_pole-param}
\end{eqnarray}
where $\displaystyle \Omega=-\sum_{i\ne G} \frac{n_i m_i^4(v_0)}{128\pi^2}$ is an uninteresting finite constant, while the functions $l_i(r)$ are
\begin{equation}
\label{eq:l_funcs}
\begin{aligned}
l_h(r=1) &=  -\frac{9\lambda^2}{16\pi^2}\Big(\frac{5\pi}{3\sqrt{3}}-3\Big),\\
l_t(r) &= \frac{3 y_t^4}{16\pi^2} \left(\frac{r}{4} - 3 + 3 F(r) A(r) \right),\\
l_W(r) &= \frac{\gL^4}{128\pi^2}\left[ 9 - r -\frac{r^2}{4}\ln r + \frac{1}{2}(r^2-12) F(r) A(r) - \frac{12 -r^2 F^2(r)}{r F(r)}A(r) \right], \\
l_Z(r) &= \frac{\gZ^4}{2 \gL^4} l_W(r),
\end{aligned}
\end{equation}
with $F(r)=\sqrt{-1+4/r}$ and $A(r) = \arctan\big(1/F(r)\big)$. 
These functions encode what remained from $\Delta \mu^2$ and $\Delta \lambda$ after using them to bring the contributions of the modes into the form seen in the last two terms of \eqref{eq:V_SM_pole-param}.
Without finite wave function renormalization ($\Delta Z=0$), the functions were computed in Ref.~\cite{Boyd:1992xn} for regularization scale $Q=M_t$. We provide a comparison of our formulae to those in Ref.~\cite{Boyd:1992xn} in App.~\ref{app:PoleBasedPot}.

The requirement of the unit residue of the Higgs propagator leads to an improved one-loop effective potential in Eq.~\eqref{eq:V_SM_pole-param} as compared to the one presented in Ref.~\cite{Boyd:1992xn}, because the function $V^{[1]}(v)$ is explicitly independent of the regularization scale.
We also mention that we obtained $V^{[1]}(v)$ through IR finite counterterms, as opposed to those appearing in Eq.~(91) of \cite{Quiros:1999jp} where the problem posed by the massless Goldstone mode was not addressed.

The Higgs curvature mass can be determined from the potential given in \eqref{eq:V_SM_pole-param}. 
It turns out that the values given in Table~\ref{tab:SMParameters} result in a Higgs curvature mass that is about 5\,\% smaller than the pole mass. 
This gives an idea on the size of the possible deviation if one parametrizes the effective potential based on the curvature mass instead of the pole mass as we do in the OPT approach in the next subsection.

\subsection{Optimized perturbation theory approach}
\label{sec:SMOPT}

In this subsection we apply the OPT procedure \cite{Chiku:1998kd} to the SM scalar potential in order to prevent the one-loop effective potential from becoming complex at small values of the background field, which occurs due to the presence of $\mu^2<0$ in the tree-level mass squared formulae of the scalars.
We do not discuss the renormalization of the OPT scheme, which can be done as in Ref.~\cite{Chiku:1998kd}.
Here, we simply drop the divergent piece $D_{\overline{\rm MS}}$ in the expression of the CW potential given in Eq.~\eqref{eq:V_SM_CW}.

Following the approach outlined in Ref.~\cite{Chiku:1998kd}, we change the classical potential by adding and subtracting a mass term in the Lagrangian
\begin{equation}
  \label{eq:SMOPT_L}
  \mathcal{L}_{\rm SM}\supset -\mu^2|\phi|^2  \longrightarrow -m^2|\phi|^2 - (\mu^2-m^2)|\phi|^2,
\end{equation}
where $m^2>0$. 
The second term is treated as an interaction term (or a finite part of the counterterm) \cite{Banerjee:1991fu} that contributes first at one-loop order. 
The new scalar potential takes the same expression as given in \eqref{eq:V_SM_cl} with $\mu^2$ changed to $m^2$. 
Therefore, within the OPT scheme we employ the replacement $\mu^2\to m^2$ in the tree-level mass squared formulae for the scalars in Eq.~\eqref{eq:SM_tree_m2}, making the Goldstone masses positive for any value of the background field $v$.

When writing the effective potential at one-loop order in the OPT scheme, the interaction term $(\mu^2-m^2)|\phi|^2$ is added to the classical potential that contains now $m^2$ instead of $\mu^2$, resulting in 
\begin{equation}
  \label{eq:SMOPT1loop}
  V_\mathrm{OPT}^{[1]}(v;\mu^2,m^2) = V_\mathrm{cl}(v;\mu^2)+V^{(1)}(v;m^2)\,.
\end{equation}
Note that the original $\mu^2$ mass parameter is restored in the classical part, but in the genuinely one-loop contributions only the new $m^2$ mass parameter appears. 
Thus, if we find a physical parametrization with $m^2>0$, then the potential will be real for all $v$.

\begin{figure}
    \centering
    \includegraphics[width=0.8\linewidth]{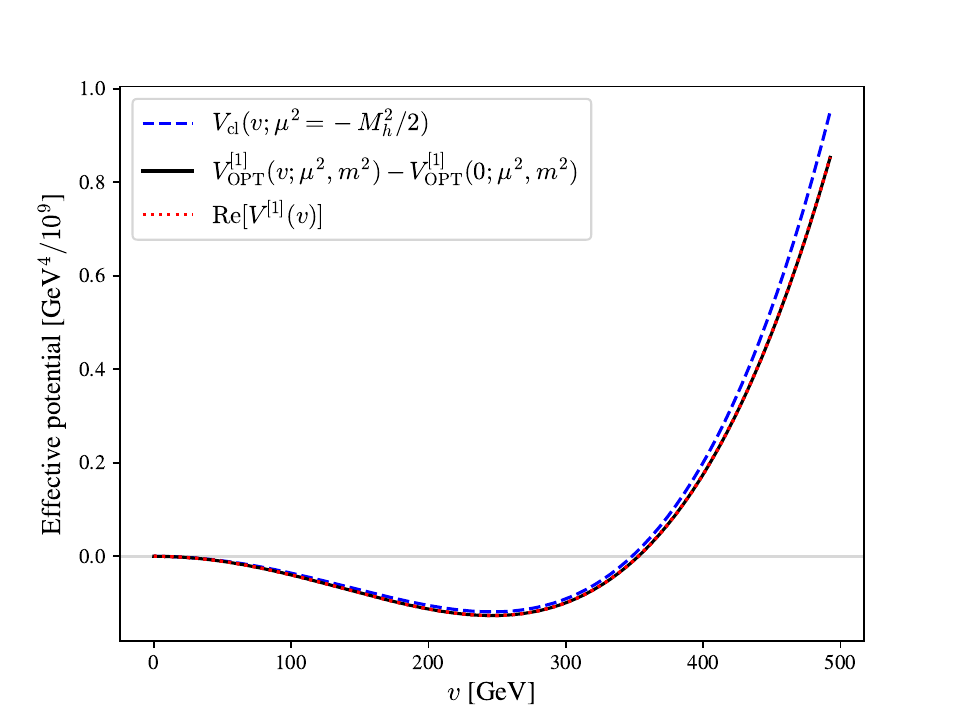}
    \caption{Comparison of SM effective potentials. 
    The OPT potential (black, solid) is compared to the real part of the potential obtained from parametrization via the pole mass (red, dotted), and the classical potential (blue, dashed).
    The OPT potential matches the one obtained from the parametrization via the pole mass to good accuracy: less than 1\,\% relative difference below the minimum.
    }
    \label{fig:SMEffPot}
\end{figure}

The one-loop effective potential has three unknown parameters: $\mu^2,~m^2,$ and $\lambda$.
In order to determine their values we need to impose three conditions on the form of the effective potential.
In the SM, the two unknown parameters $\mu^2$ and $\lambda$ are determined by fixing the value of the VEV (minimum of the potential) and the value of the Higgs mass:
\begin{align}
    &\text{Condition 1:}\quad \left.\frac{\partial V_\mathrm{OPT}^{[1]}(v;\mu^2,m^2)}{\partial v}\right|_{v=v_0} = 0\,, \\
    &\text{Condition 2:}\quad \left.\frac{\partial^2 V_\mathrm{OPT}^{[1]}(v;\mu^2,m^2)}{\partial v^2}\right|_{v=v_0} = M_h^2\,.
\end{align}
Here we used the approximation that the Higgs pole and curvature masses are equal.
In  Sec.~\ref{sec:PoleMassparametrization} we saw that the latter is smaller by about 5\,\%.
For the VEV and Higgs boson mass we take the values given in Table~\ref{tab:SMParameters}.
To fix the third parameter, we employ the principle of minimum sensitivity (PMS), namely that the potential at the minimum (physical point) should not depend on the value of $m^2$:
\begin{equation}
    \text{Condition 3:}\quad \left.\frac{\partial V_\mathrm{OPT}^{[1]}(v;\mu^2,m^2)}{\partial m^2}\right|_{v=v_0}=0\,.
\end{equation}
Choosing $Q=M_t$ for the regularization scale, with the value of $M_t$ given in Table~\ref{tab:SMParameters}, the solution of the three conditions is
\begin{equation}
    \label{eq:OPTSMsolutions}
    m^2=69~094\GeV^2,\quad\lambda=0.12~861,\quad\mu^2=-8~847.9\GeV^2\,.
\end{equation}
Since $m^2>0$, it is guaranteed that the OPT one-loop effective potential is real for all values of $v$, as shown in Fig.~\ref{fig:SMEffPot}. 
As a comparison we have also plotted the classical scalar potential and the real part of the potential in Eq.~\eqref{eq:V_SM_pole-param} obtained in the previous subsection.

\section{Effective potential in the singlet extension\label{sec:sSM_pot}}

Scalar extensions of the SM have been popular in the past decades \cite{Sher:1988mj,Branco:2011iw,Barger:2008jx,Barger:2007im} as they provide an interesting phenomenology from a particle physics as well as from a cosmological point of view. 
The differences between particular models lie in the number of new scalars introduced, as well as in their transformation properties. 
Two very well studied models are the two Higgs doublet model (2HDM) \cite{Lee:1973iz} and the scalar singlet extension of the SM \cite{Barger:2008jx,Chiang:2017nmu}. 
In this paper we are going to investigate the one-loop effective potential of the latter.

\subsection{Classical potential}
\label{sec:TheClassicalPotential}

We consider the model where we add a single complex scalar field $\chi$ to the SM:
\begin{equation}
    \label{eq:chi_field}
    \chi(x)=\frac{1}{\sqrt{2}}\big(w + s(x) + \ri\chi_2(x)\big)\,.
\end{equation}
Similarly to the BEH doublet, we allow the singlet scalar to have a non-vanishing vacuum expectation value $w$.
The fluctuations around the constant background are described by the scalar fields $s(x)$ and $\chi_2(x)$ (the new Goldstone boson).

To allow for non-trivial phenomenology (i.e.~having a not completely decoupled scalar), we allow mixing between the BEH doublet $\phi$ and the singlet $\chi$, and we consider the potential
\begin{equation}
    \label{eq:SingletPotential}
    \mathcal{V}(\phi,\chi)=\mu_\phi^2|\phi|^2 + \lambda_\phi|\phi|^4 +\mu_\chi^2|\chi|^2 + \lambda_\chi|\chi|^4 + \lambda'|\phi|^2|\chi|^2\,,
\end{equation}
from which the classical scalar potential is obtained via averaging over all fluctuating fields:
\begin{equation}
    \label{eq:SingletClassicalPotential}
    V_\mathrm{cl}(v,w;\mu_\phi^2,\mu_\chi^2)= \frac{\mu_\phi^2}{2}v^2 + \frac{\lambda_\phi}{4}v^4 + \frac{\mu_\chi^2}{2}w^2 + \frac{\lambda_\chi}{4}w^4 + \frac{\lambda'}{4}v^2w^2\,.
\end{equation}
As the classical potential is now a function of two background fields, the conditions for having a potential that is bounded from below, and has a non-trivial minimum are more involved. 
For $\lambda_{\phi,\chi}>0$ the requirement is \cite{Senaha:2020mop}
\begin{equation}
    \label{eq:BoundednessCondition}
    \lambda'\geq -2\sqrt{\lambda_\chi\lambda_\phi}\,.
\end{equation}
The boundedness condition in itself does not constrain positive values of $\lambda'$.

The background-dependent Goldstone masses are obtained as
\begin{subequations}
\label{eq:GoldstoneMass}
\begin{align}
    \label{eq:GoldstonePhiMass}
    m_{\phi_i}^2(v,w;\mu_\phi^2)&=\frac{1}{v}\frac{\partial V_\mathrm{cl}(v,w;\mu_\phi^2,\mu_\chi^2)}{\partial v}=\mu_\phi^2 + \lambda_\phi v^2 + \frac{\lambda'}{2}w^2\,,\quad \text{for }i=2,3,4\\
    \label{eq:GoldstoneChiMass}
    m_{\chi_2}^2(v,w;\mu_\chi^2)&=\frac{1}{w}\frac{\partial V_\mathrm{cl}(v,w;\mu_\phi^2,\mu_\chi^2)}{\partial w}=\mu_\chi^2 + \lambda_\chi w^2 + \frac{\lambda'}{2}v^2\,.
\end{align}
\end{subequations}
Due to the mixing of scalar fields, the states $h(x)$ and $s(x)$ are not mass-eigenstates. 
The curvature masses for the Higgs boson and the singlet scalar are obtained by diagonalizing the Hessian of the classical potential:
\begin{equation}
    \label{eq:TreelevelMassMatrix}
    \mathbf{M}_\mathrm{cl}^2(v,w;\mu_\phi^2,\mu_\chi^2)= 
    \begin{pmatrix}
        \mu_\phi^2+3\lambda_\phi v^2+\frac{\lambda'}{2}w^2 & \lambda'vw \\
        \lambda'vw & \mu_\chi^2+3\lambda_\chi w^2+\frac{\lambda'}{2}v^2
    \end{pmatrix}\,.
\end{equation}
The general formulae for the eigenvalues $m^2_{\mathrm{cl},\pm}(v,w;\mu_\phi^2,\mu_\chi^2)$ are complicated, but they can be simplified at the minimum $(v_0,w_0)$ of the potential.
In this case the Goldstone masses in Eq.~\eqref{eq:GoldstoneMass} vanish and we can eliminate the $\mu^2_{\phi,\chi}$ parameters in favour of the vacuum expectation values:
\begin{equation}
    \label{eq:muvevrelation}
    \mu_\phi^2 = -\lambda_\phi v_0^2 - \frac{\lambda'}{2}w_0^2\,,\quad \mu_\chi^2= -\lambda_\chi w_0^2 - \frac{\lambda'}{2}v_0^2\,.
\end{equation}
Exploiting the relations in Eq.~\eqref{eq:muvevrelation}, the classical mass squared eigenvalues of Eq.~\eqref{eq:TreelevelMassMatrix} in the minimum of the potential are
\begin{equation}
\label{eq:TreeLevelScalarMasses}
    m_{\mathrm{cl},\pm}^2\Big(v_0,w_0;\mu_\phi^2,\mu_\chi^2\Big) = \lambda_\phi v_0^2 + \lambda_\chi w_0^2 \pm \sqrt{\left(\lambda_\phi v_0^2-\lambda_\chi w_0^2\right)^2+(\lambda'v_0w_0)^2}\,.
\end{equation}
Requiring that both classical scalar masses are real leads to a constraint on the mixing $\lambda'$:
\begin{equation}
    \label{eq:RealMassCondition}
    |\lambda'|\leq 2\sqrt{\lambda_\chi\lambda_\phi}\,,
\end{equation}
which extends the already existing boundedness condition of Eq.~\eqref{eq:BoundednessCondition} to positive values of $\lambda'$ as well.

In our investigation of the singlet extension of the SM we set the smaller eigenvalue to be equal to the observed Higgs mass squared, and the larger one to be the yet unknown mass of the singlet:
\begin{equation}
    \label{eq:MassSet}
    m_{\mathrm{cl},-}^2\Big(v_0,w_0;\mu_\phi^2,\mu_\chi^2\Big)=M_h^2\,,\quad m_{\mathrm{cl},+}^2\Big(v_0,w_0;\mu_\phi^2,\mu_\chi^2\Big)=M_s^2\,.
\end{equation}
Experimental constraints exclude the possiblity of the opposite assignment \cite{Peli:2022ybi,Robens:2015gla}. The expressions for the masses in Eqs.~\eqref{eq:TreeLevelScalarMasses} and \eqref{eq:MassSet} are satisfied by two separate parameter points $\{\lambda_\phi,\lambda_\chi\}$ for any physical $\lambda'$.
This double covering of the same physical observables is removed by requiring that the Higgs mass is dominantly due to the vacuum expectation value of the $\phi$ field.
This requirement is satisfied when
\begin{equation}
    \label{eq:PhysicalSelection}
    \lambda_\phi v_0^2 \leq \lambda_\chi w_0^2\,, \text{ for any }\lambda'\in \Big[-\lambda'_{\max},\lambda'_{\max}\Big].
\end{equation}
Indeed, at $\lambda'=0$ the lighter mass is then given by $m_{\mathrm{cl},-}^2=2\lambda_\phi v_0^2$ as in the SM.
The limiting value $\lambda'_{\max}$ is determined for a given set $\{M_s,w_0\}$ by the condition $\lambda_\phi v_0^2 = \lambda_\chi w_0^2$, which gives
\begin{equation}
    \label{eq:LambdahsMax}
    \lambda'_{\max}=\frac{M_s^2-M_h^2}{2v_0w_0},\text{ for } M_s^2>M_h^2\,.
\end{equation}
There are no solutions to the parametrization if $|\lambda'|>\lambda'_{\max}$.
This constraint on the mixing $\lambda'$ is stricter than that of Eq.~\eqref{eq:RealMassCondition}, thus any parametrization in terms of non-negative squared scalar masses that results in real parameter values will yield a real potential bounded from below.

\subsection{One-loop corrections}

The tree-level treatment described in the previous subsection is general, that is, it holds for any theory in which the extension includes a singlet scalar.
However, the corrections to it incorporate loops involving all fields in the model.
We describe the general approach to the one-loop parametrization in this subsection, and present the formulae obtained in the minimal singlet extension and in a more complete model in the following section.

The singlet extended scalar potential in Eq.~\eqref{eq:SingletPotential} has 5 unknown parameters: $\mu_{\phi}^2$, $\mu_{\chi}^2$, $\lambda_\phi$, $\lambda_\chi,$ and $\lambda'$. 
In order to fix them, we extend the procedure applied to the SM in Sec.~\ref{sec:SMOPT}. 
In addition to the known mass $M_h$ of the Higgs boson and the vacuum expectation value $v_0$ of the $\phi$ field, we pick a fixed value for the mass $M_s$ of the $s$ scalar and another one for the vacuum expectation value $w_0$ of the $\chi$ field. 
These give 4 conditions for the 5 parameters of the potential. 
We investigate the parametrization by treating $\lambda'$ as a free parameter.

In general the one-loop effective potential is
\begin{equation}
    V^{[1]}_\mathrm{eff}(v,w;\mu_\phi^2,\mu_\chi^2)=V_\mathrm{cl}(v,w;\mu_\phi^2,\mu_\chi^2)+\sum_{i}n_iV_{\mathrm{CW}}\big(m_i^2(v,w;\mu_\phi^2,\mu_\chi^2)\big)\,,
\end{equation}
where the sum is over all fields present in the given model.
The background-dependent masses in the one-loop correction depend in general on both vacuum expectation values.

The singlet extended potential suffers from the same problem as that in the SM, it becomes complex in a certain region of $\{v,w\}$.
To deal with this problem, we proceed as in Sec.~\ref{sec:SMOPT} and introduce new mass parameters for both scalars (c.f. Eq.~\eqref{eq:SMOPT_L}) denoted with $m_\phi^2$ and $m_\chi^2$:\begin{equation}
  \mathcal{L}\supset -\mu_\phi^2|\phi|^2 -\mu_\chi^2|\chi|^2 \longrightarrow -m_\phi^2|\phi|^2 - m_\chi^2|\chi|^2 - \Big[(\mu_\phi^2-m_\phi^2)|\phi|^2 + (\mu_\chi^2-m_\chi^2)|\chi|^2\Big]\,.
\end{equation}
The terms in square brackets are treated as interactions, first appearing in a one-loop level calculation of the effective potential. 
These terms restore the original mass parameters $\mu_{\phi/\chi}^2$ in the classical part of the one-loop level OPT effective potential, as in Eq.~\eqref{eq:SMOPT1loop}, while the genuinely one-loop part added to this contains the new mass parameters $m_{\phi/\chi}^2$:
\begin{equation}
    \label{eq:1loopSingletEffPotOPT}
    V_\mathrm{OPT}^{[1]}(v,w;\mu_\phi^2,\mu_\chi^2,m_\phi^2,m_\chi^2) = V_\mathrm{cl}(v,w;\mu_\phi^2,\mu_\chi^2) + \sum_i n_iV_{\mathrm{CW}}\big(m_i^2(v,w;m_\phi^2,m_\chi^2)\big)\,.
\end{equation}

The conditions to fix the parameters in the effective potential Eq.~\eqref{eq:1loopSingletEffPotOPT} are similar to those used in the SM.
First of all, $v_0$ and $w_0$ are imposed as minima of the bivariate potential:
\begin{align}
    \text{Condition 1:}\quad&\left.\frac{\partial V_\mathrm{OPT}^{[1]}(v,w;\mu_\phi^2,\mu_\chi^2,m_\phi^2,m_\chi^2)}{\partial v}\right|_{v=v_0,w=w_0} = 0\,, \\
    \text{Condition 2:}\quad&\left.\frac{\partial V_\mathrm{OPT}^{[1]}(v,w;\mu_\phi^2,\mu_\chi^2,m_\phi^2,m_\chi^2)}{\partial w}\right|_{v=v_0,w=w_0} = 0 \,.
\end{align}
Secondly, we require that the curvature masses at the minimum of the potential are equal to the Higgs boson mass $M_h$ and the new scalar mass $M_s$.
Due to mixing between the physical states, one has to look at the eigenvalues of the Hessian 
\begin{equation}
    \mathbf{M}^2(v,w;\mu_\phi^2,\mu_\chi^2,m_\phi^2,m_\chi^2)=
    \begin{pmatrix}
        \partial_v^2 V_\mathrm{OPT}^{[1]} & \partial_v\partial_w V_\mathrm{OPT}^{[1]} \\
        \partial_w\partial_v V_\mathrm{OPT}^{[1]} & \partial_w^2 V_\mathrm{OPT}^{[1]}
    \end{pmatrix},
\end{equation}
and relate those to the masses. Let these eigenvalues be given by $m_\pm^2(v,w;\mu_\phi^2,\mu_\chi^2,m_\phi^2,m_\chi^2)$, with $m_+ > m_->0$.
We assumed that the Higgs boson is the lighter scalar, thus the conditions for the masses are:
\begin{align}
    \text{Condition 3:}\quad&m_-^2(v_0,w_0;\mu_\phi^2,\mu_\chi^2,m_\phi^2,m_\chi^2) = M_h^2\,, \\
    \text{Condition 4:}\quad&m_+^2(v_0,w_0;\mu_\phi^2,\mu_\chi^2,m_\phi^2,m_\chi^2) = M_s^2\,.
\end{align}
Thirdly, we have two stationarity conditions based on the PMS,
\begin{align}
    \text{Condition 5:}\quad&\left.\frac{\partial V_\mathrm{OPT}^{[1]}(v,w;\mu_\phi^2,\mu_\chi^2,m_\phi^2,m_\chi^2)}{\partial m_\phi^2}\right|_{v=v_0,w=w_0}=0\,, \\
    \text{Condition 6:}\quad&\left.\frac{\partial V_\mathrm{OPT}^{[1]}(v,w;\mu_\phi^2,\mu_\chi^2,m_\phi^2,m_\chi^2)}{\partial m_\chi^2}\right|_{v=v_0,w=w_0}=0\,.
\end{align}
We use these six conditions to fix the values of $m_{\phi/\chi}^2$, $\mu_{\phi/\chi}^2$ and $\lambda_{\phi/\chi}$, while $\lambda'$ is kept as a free parameter.

For a given model, the existence of a real solution to the parametrization conditions is not guaranteed due to the presence of logarithmic contributions involved \cite{Marko:2015gpa}.
While an estimate for the region of validity can be given in any model based on tree-level considerations, in general the loop contributions may be significant enough that the parameter regions where a solution exists are only available through trial and error.

\section{Parametrization of SM extensions with a singlet scalar \label{sec:param_sSM_pot}}

In this section we are going to showcase the parametrization outlined in the previous section in two models.
In both cases the potential $\mathcal{V}(\phi,\chi)$ in the Lagrangian is the same, given in Eq.~\eqref{eq:SingletPotential}.
First, we are going to present the parametrization of a model that is the SM plus a complex singlet scalar only.
Second, we take a look at the so-called {\it super-weak extension of the standard model} that includes a complex singlet scalar among other non-scalar new degrees of freedom.

\subsection{SM with a singlet scalar}
\label{sec:SingletExtendedSM}

This model involves no new fields apart from $\chi(x)$, parametrized as in Eq.~\eqref{eq:chi_field}, thus the loop corrections to the effective potential will be the same as those in the SM, plus the contribution of the scalar $s(x)$ and the Goldstone $\chi_2(x)$.
In this model the background-dependent tree-level masses of the gauge bosons and the top quark remains those given in Eqs.~\eqref{eq:SMmassesWZ}-\eqref{eq:SMmassesTop}.
All Goldstone and scalar masses are modified due to the mixing.
The Goldstone masses were given in Eqs.~\eqref{eq:GoldstonePhiMass}-\eqref{eq:GoldstoneChiMass}, while the scalar masses are understood as the eigenvalues of the Hessian in Eq.~\eqref{eq:TreelevelMassMatrix}.

\begin{figure}[t]
    \centering
    \includegraphics[width=\linewidth]{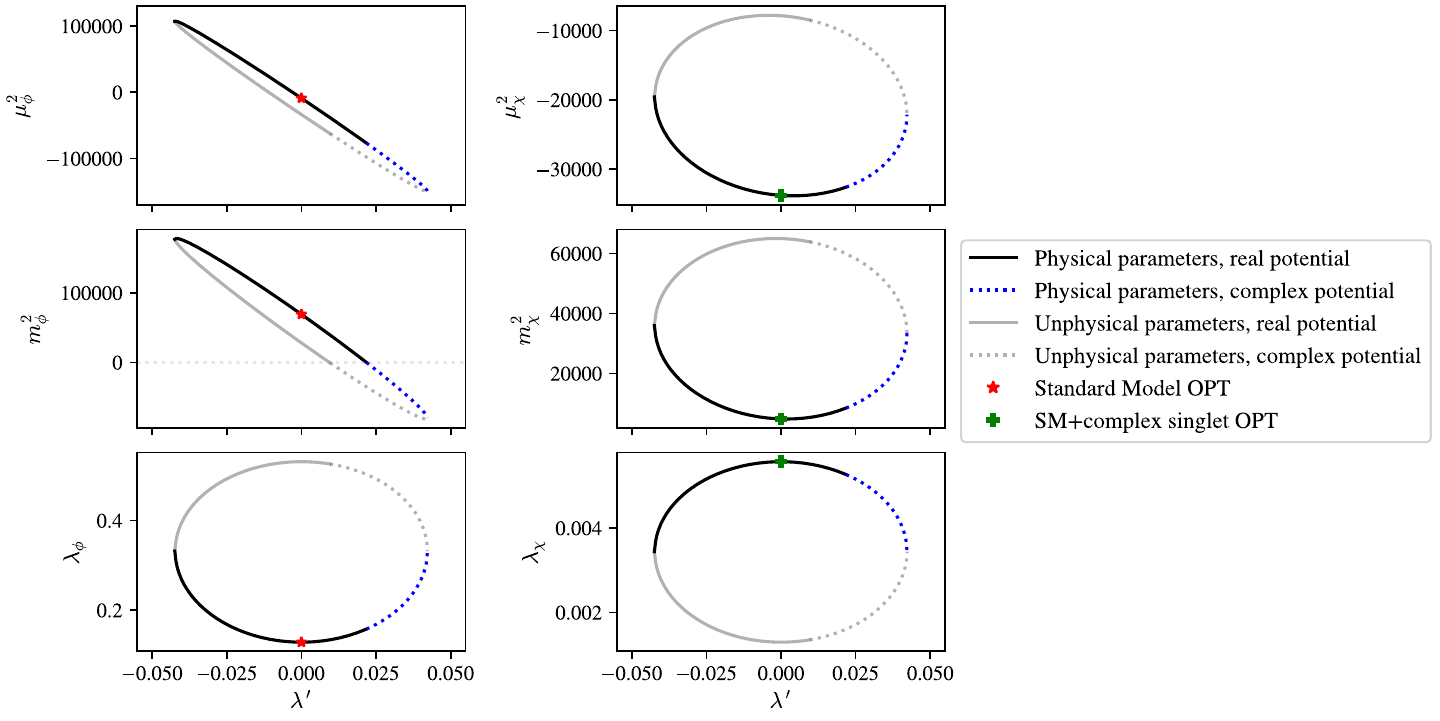}
    \caption{Parametrization of the singlet extended SM for $M_s=260\GeV$, $w_0=10v_0$, and regularization scale $Q=M_t$.
    The circular shapes indicate the possible real solutions to the parametrization conditions.
    The unphysical solutions are shown in light gray.
    The blue dotted part of the physical parameters corresponds to solutions where the effective potential becomes complex for some region satisfying $v<v_0$ and $w<w_0$, but the minimum remains real.
    Additionally, at $\lambda'=0$ the SM sector and the singlet are completely decoupled, and they can be parametrized separately (as in Sec.~\ref{sec:SMOPT}), leading to the points shown with red stars and green crosses.}
    \label{fig:260GeVSingletParameters}
\end{figure}

We show an example parametrization in Fig.~\ref{fig:260GeVSingletParameters} with a scalar mass $M_s=260\GeV$, $\chi(x)$ vacuum expectation value $w_0=10v_0$, and regularization scale $Q=M_t$.
The figure shows all 6 fitted parameters as functions of the quartic mixing $\lambda'$.
Solutions exist for a narrow range of $\lambda'$, symmetric around $\lambda'=0$ in rough agreement with the conclusion we drew in Eq.~\eqref{eq:LambdahsMax} from a tree-level analysis.
The characteristic feature of the figure is the elliptic shape of all parameter curves: there are two solutions at any value of $\lambda'$ that can be related to each other by a simple reflection.
As discussed around Eq.~\eqref{eq:PhysicalSelection}, this doubling is expected and we can select the physical solution as that with the lighter scalar mass (i.e.,~the Higgs boson) mostly given by the vacuum expectation value of the $\phi$ field.
As the scalar sectors completely decouple at $\lambda'=0$, the left-hand side plots in Fig.~\ref{fig:260GeVSingletParameters} should reproduce the SM result in Eq.~\eqref{eq:OPTSMsolutions} (shown with red stars).
The physical selection in Eq.~\eqref{eq:PhysicalSelection} is equivalent to choosing the half of the elliptic curve on which the SM point lies.

The reason for parametrizing the model using the OPT approach was to
obtain an effective potential that is real for any value of $v,w$.
In the SM this was shown to be possible in Sec.~\ref{sec:SMOPT}, however the SM is a theory with enough observables to uniquely fix all of its parameters.
Any extension to the SM will have new degrees of freedom, thus yet unobserved values for certain physical quantities.
However, there is no guarantee in the parametrization conditions that the potential will remain fully real for any solution.
In Fig.~\ref{fig:260GeVSingletParameters} the parameter value $m_\phi^2$ becomes negative above\footnote{The tilt of the elliptical parameter curves of $\mu_{\phi/\chi}^2$ and $m_{\phi/\chi}^2$ is negative if the new scalar is heavier than the Higgs, which is our choice.} a certain value of $\lambda'$ indicated by the dotted blue line.
At these parameter values, the potential is complex in a given region centered at the origin of $\{v,w\}$, however it is real at the minimum.

The parametrization depends on the values of $M_s$ and $w_0$.
For a particular choice, a generic solution is shown in Fig.~\ref{fig:260GeVSingletParameters}.
In fact, a real parametrization does not always exist.
The characteristic features of the parametrization can be understood mostly using tree-level relations detailed in Sec.~\ref{sec:TheClassicalPotential}.
In what follows we explore the effects of changing the (mostly) arbitrary input values for $M_s$ and $w_0$.
From Eq.~\eqref{eq:LambdahsMax} we determine that the non-trivial sector for these parameters is when $M_s$ and $M_h$ do not coincide, and $w_0$ is not much larger than the scales set by the masses.

\begin{figure}
    \centering
    \includegraphics[width=\linewidth]{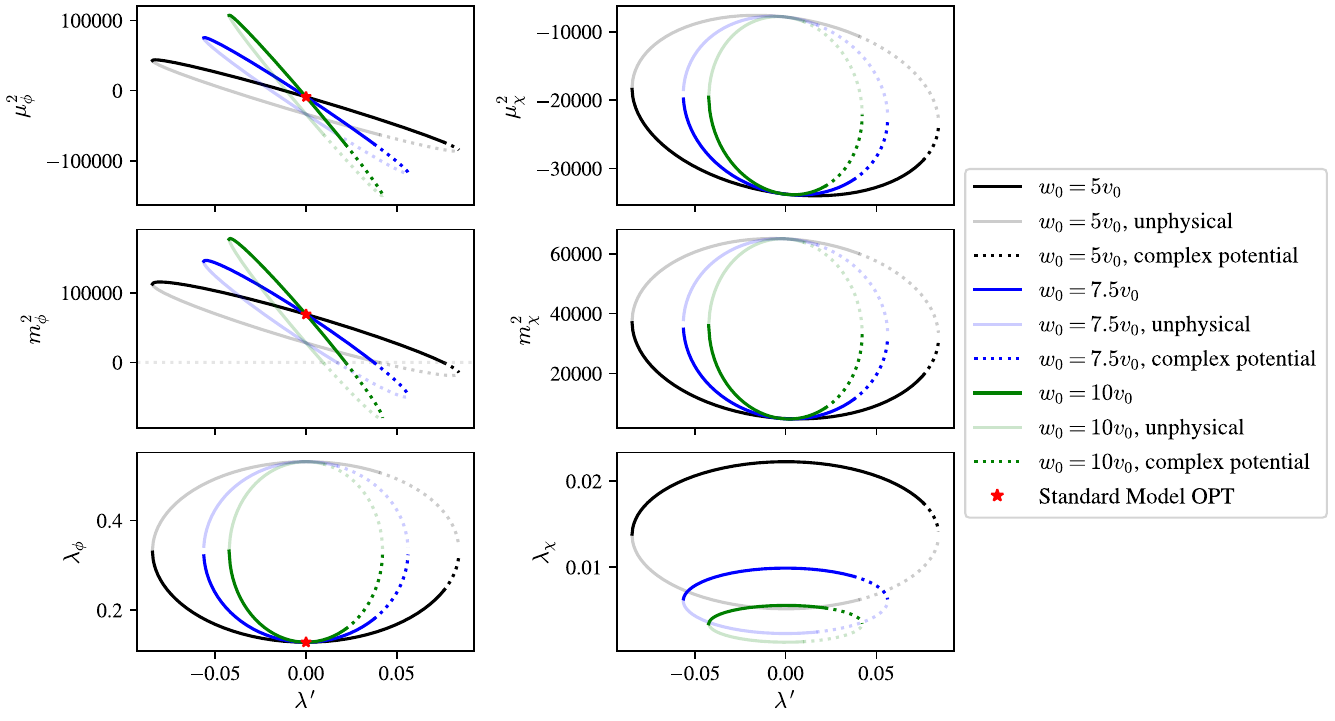}
    \caption{Parametrization of the singlet extended SM for $M_s=260\GeV$ at various values of VEV $w_0$. The solid curves indicate physical solutions where the potential remains real for all values of the background fields. The light solid curves are the unphysical solutions, while the dotted lines indicate parameter values where the potential becomes complex in some region of $v<v_0$ and $w<w_0$.}
    \label{fig:VaryingVEV}
\end{figure}

Changing the vacuum expectation value $w_0$ while keeping the mass $M_s$ fixed will result in parameter ranges shown in Fig.~\ref{fig:VaryingVEV}.
In all plots we see the changing of the physical $\lambda'$ domain, as described by Eq.~\eqref{eq:LambdahsMax}.
For $\mu_\phi^2$ and $m_\phi^2$ (and less obviously for $\mu_\chi^2$ and $m_\chi^2$) we see a change in the tilt of the elliptical curves with changing VEVs.
From tree-level relations one can find the difference between the values of $\mu_{\phi/\chi}^2$ for the two extreme values for $\lambda'=\pm \lambda'_{\max}$,
\begin{gather}
    \left.\mu_\phi^2\right|_{\lambda'=-\lambda'_{\max}} - \left.\mu_\phi^2\right|_{\lambda'=\lambda'_{\max}} \simeq \frac{M_s^2-M_h^2}{2}\frac{w_0}{v_0}\,, \\
    \left.\mu_\chi^2\right|_{\lambda'=-\lambda'_{\max}} - \left.\mu_\chi^2\right|_{\lambda'=\lambda'_{\max}} \simeq \frac{M_s^2-M_h^2}{2}\frac{v_0}{w_0}\,.
\end{gather}
As a base assumption we have $M_s>M_h$, thus for an increasing $w_0$ we expect the solution curves for $\mu_\phi^2$ and $\mu_\chi^2$ to rotate clockwise, which is confirmed by Fig.~\ref{fig:VaryingVEV}.

The solutions for $\lambda_\chi$ in Fig.~\ref{fig:VaryingVEV} shift upwards with decreasing vacuum expectation values $w_0$. 
The border between the physical and unphysical solutions at tree-level satisfies the simple relation $\lambda_\phi v_0^2=\lambda_\chi w_0^2$ (see Eq.~\eqref{eq:PhysicalSelection}).
In particular, the minimum value of the physical $\lambda_\chi$ is expressed as
\begin{equation}
    \lambda_\chi^\mathrm{(min)}=\frac{M_s^2+M_h^2}{4w_0^2}\,,
\end{equation}
which describes the effect we see in the last plot of Fig.~\ref{fig:VaryingVEV}.

\begin{figure}
    \centering
    \includegraphics[width=\linewidth]{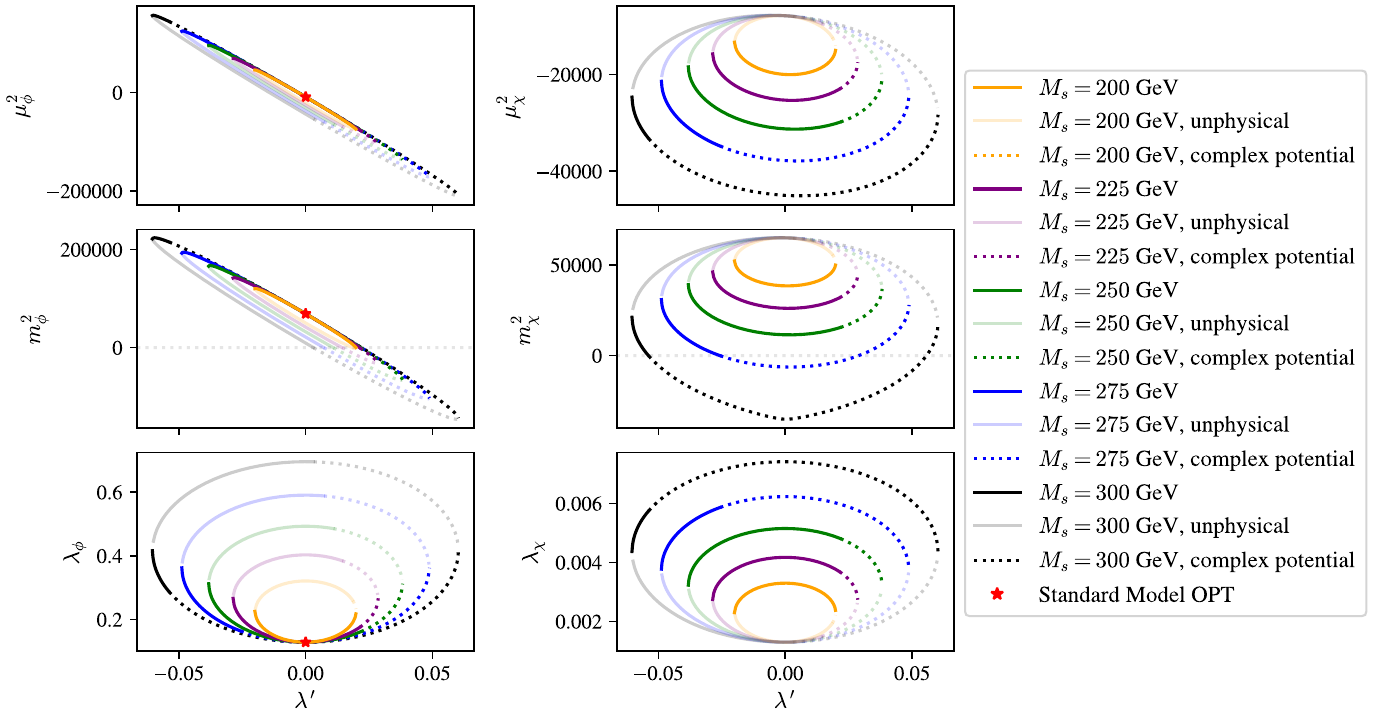}
    \caption{Parametrization of the singlet extended SM for $w_0=10v_0$ at various values of the scalar mass $M_s$.
    The solid curves indicate physical solutions where the potential remains real for all values of the background fields.
    The light solid curves are the unphysical solutions, while the dotted lines indicate physical parameter values where the potential becomes complex in some region of $v<v_0$ and $w<w_0$.}
    \label{fig:VaryScalarMass}
\end{figure}

Next, we look at the parametrization of the singlet scalar extended SM for a fixed VEV $w_0$ with varying the $M_s$ scalar mass.
The resulting parameters are shown in Fig.~\ref{fig:VaryScalarMass} for $w_0=10v_0$ and singlet scalar masses in the domain $M_s\in [200,\,300]~$GeV.
We see an interesting new feature when $M_s$ is varied: $m_\chi^2$ can become negative for a symmetric region around $\lambda'=0$, as indicated by the dotted lines in Fig.~\ref{fig:VaryScalarMass}.
This is not necessarily an issue, as the minimum point of the potential is still real (i.e., the parametrization exists).
The largest singlet scalar mass for which the potential remains real at $\lambda'=0$ (i.e., at the completely decoupled scalar sector limit) is approximated by setting the values of $\mu_\chi^2$ and $\lambda_\chi$ to their tree-level expressions and solving the PMS condition,
\begin{equation}
    \frac{\partial V_{\mathrm{CW}}\big(m_\text{cl,+}^2(v,w;m_\phi^2,m_\chi^2)\big)}{\partial m_\chi^2}+\frac{\partial V_\mathrm{CW}\big(m_{\chi_2}^2(v,w;m_\phi^2,m_\chi^2)\big)}{\partial m_\chi^2}=0
\end{equation}
to obtain $m_\chi^2$.
The upper bound for having a real solution is independent of the vacuum expectation value $w_0$, and we find the condition for having a real potential at $\lambda'=0$ to be
\begin{equation}
    \label{eq:MassBound}
    \frac{M_s}{Q} \leq \frac{\sqrt{2\mathrm{e}}}{3^{3/8}}\,.
\end{equation}
Larger $M_s$ values are still compatible with the parametrization, but the resulting potentials will become complex for an increasing range in $\lambda'$.
The parametrization breaks down (i.e., we would require complex parameters to satisfy the conditions at the minimum of the potential) for $M_s/Q\gtrsim 1.74$.

\begin{figure}
    \centering
    \includegraphics[width=\linewidth]{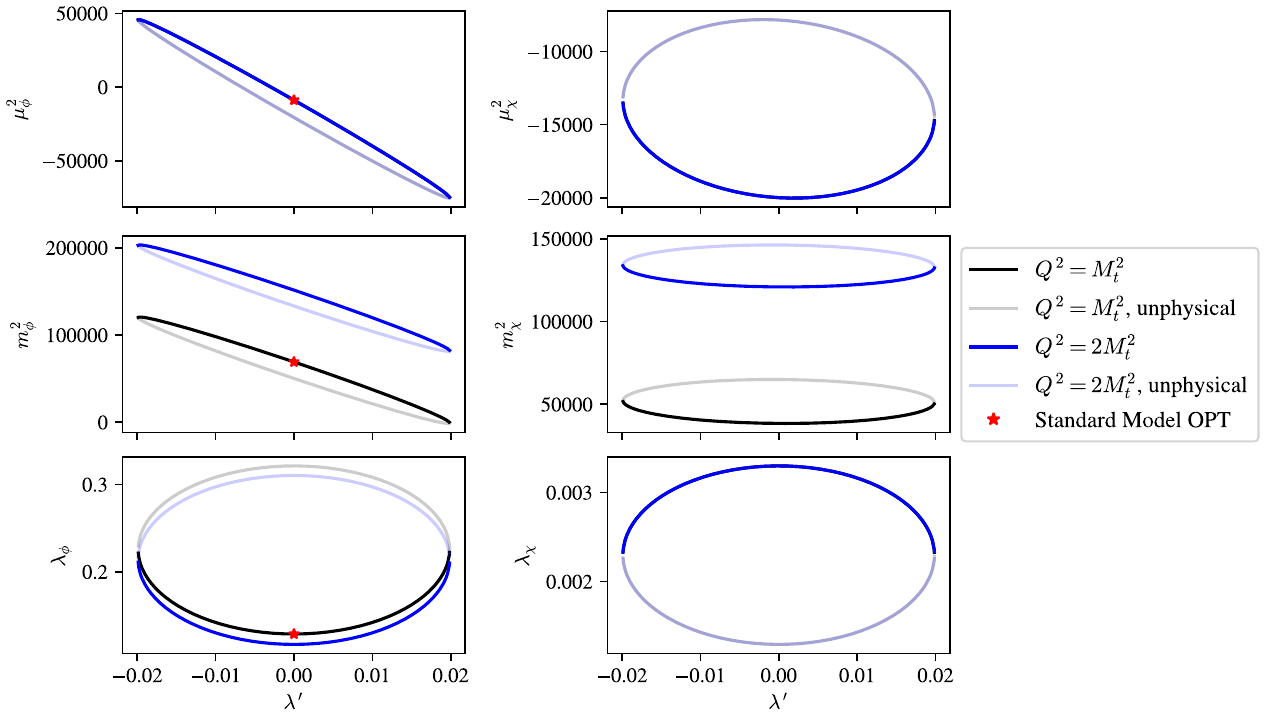}
    \caption{Comparison of parametrizations of the singlet extended SM with $w_0=10v_0$ and $M_s=200\GeV$ for two values of the regularization scale $Q$.
    Larger regularization scale leads to larger values of the $m^2$ parameters.
    Positivity of $m^2$ is required for a fully real effective potential, and larger $Q$ can accommodate heavier scalar masses $M_s$ as given by the bound Eq.~\eqref{eq:MassBound}.}
    \label{fig:SingletQ2}
\end{figure}

The choice of the regularization scale $Q$ influences the existence of real solutions to the parametrization, but their actual value is unphysical.
From Eq.~\eqref{eq:MassBound} we see that larger values for the unphysical scale $Q$ can accommodate heavier singlet scalars while the potential remains real.
In Fig.~\ref{fig:SingletQ2} we compare two parametrizations done with different regularization scales, $Q=M_t$ and $Q=\sqrt{2}M_t$.
The value of the quartic coupling depends logarithmically on the regularization scale, which is a standard result, while the mass squared parameter $\mu^2_{\phi/\chi}$ is independent of it.
The shifted mass parameters $m^2_{\phi/\chi}$ introduced in the OPT scheme only appear in the one-loop part of the effective potential, thus their values depend heavily on the regularization scale.
It is seen in Fig.~\ref{fig:SingletQ2} that a parametrization done at a larger $Q$ results in larger positive mass squared parameters $m^2_{\phi/\chi}$, thus a real potential even for heavier scalars, in accordance with Eq.~\eqref{eq:MassBound}.

\subsection{Super-weak model}

In this section we apply the OPT formalism to a model which includes extra massive fermionic degrees of freedom.
The model chosen here is the SWSM \cite{Trocsanyi:2018bkm}.
This model is an economic U(1)$_z$ extension of the SM  focusing on experimentally reachable light new physics, and probing the parameter space for various BSM scenarios.
The spectrum of the model extends that of the SM by a complex singlet scalar field $\chi$, 3 generations of right-handed neutrinos $N_i$ (RHN), and a massive gauge boson of the new U(1)$_z$ symmetry $Z'$.
The U(1)$_z$ symmetry is spontaneously broken by the non-zero VEV of $\chi$, generating mass for $Z'$ and the RHNs through the Higgs mechanism.
In Refs.~\cite{Iwamoto:2021fup,Seller:2022noz} it was shown that the model is capable of explaining MeV scale sterile neutrino dark matter if the coupling $g_z\sim\mathcal{O}(10^{-5})$ is tiny, and the VEV of $\chi$ is comparable to the SM VEV, $2v_0 < w_0 < 100v_0$.
We use this region of the parameter space in the following.

The effective potential depends mostly on the heavy particles ($m_\mathrm{heavy}\gtrsim M_W$) in the spectrum.
For the parameter space outlined above $Z'$ and the DM candidate RHN $N_1$ are light, and their small contribution to the effective potential is neglected.
The masses of the heavier neutrinos only depend on the VEV of the $\chi$ field (SM contribution is neglected due to the see-saw mechanism):
\begin{equation}
    \label{eq:RHNmass}
    m_{N_{j}}^2(w)\approx\frac{y_{N_{j}}^2w^2}{2}\,,\text{ for }j=2,3\,.
\end{equation}
Each RHN carries $n_{N_j}=2$ degree of freedom (particle and antiparticle), and the Yukawa couplings $y_{N_j}$ are free parameters of the model.
The masses of the remaining SM degrees of freedom remain approximately unchanged: the gauge boson masses were given in Eq.~\eqref{eq:SMmassesWZ} (an $\mathcal{O}\big(g_z^2/\gZ^2\big)$ correction to the $Z$ boson mass is neglected), and the top quark mass is \eqref{eq:SMmassesTop}. 
The scalar sector is the same as that detailed in Sec.~\ref{sec:TheClassicalPotential}, in particular see Eq.~\eqref{eq:GoldstoneMass} and Eq.~\eqref{eq:TreelevelMassMatrix} for the Goldstone and scalar masses.

The addition of RHNs to the spectrum only affects the singlet direction in the effective potential, as their masses only depend on $w$.
Additional fermions decrease the one-loop value of the mass squared parameter, but increase that of the quartic coupling.
Assuming that in the SWSM the two heavy RHNs have approximately the same mass, the one-loop corrections to the parameters at $\lambda'=0$ are approximately (neglecting all contributions apart from the RHNs)
\begin{equation}
\begin{split}
    \label{eq:SWSMparametershift}
    \left.\mu_\chi^2\right|_{\lambda'=0} - \left.\mu_{\chi,\text{tree}}^2\right|_{\lambda'=0}&\approx -\frac{y_N^4w_0^2}{16\pi^2}\,,\\
    \left.\lambda_\chi\right|_{\lambda'=0} - \left.\lambda_{\chi,\text{tree}}\right|_{\lambda'=0}&\approx \frac{y_N^4}{2w_0^2}\log\left(\frac{y_N^2w_0^2}{2Q^2}\right)\,.
\end{split}
\end{equation}
As $w_0$ is constrained to be larger than the BEH VEV, the contribution of RHNs is largely negligible unless their masses lie well above the electroweak scale.

\begin{figure}[t]
    \centering
    \includegraphics[width=\linewidth]{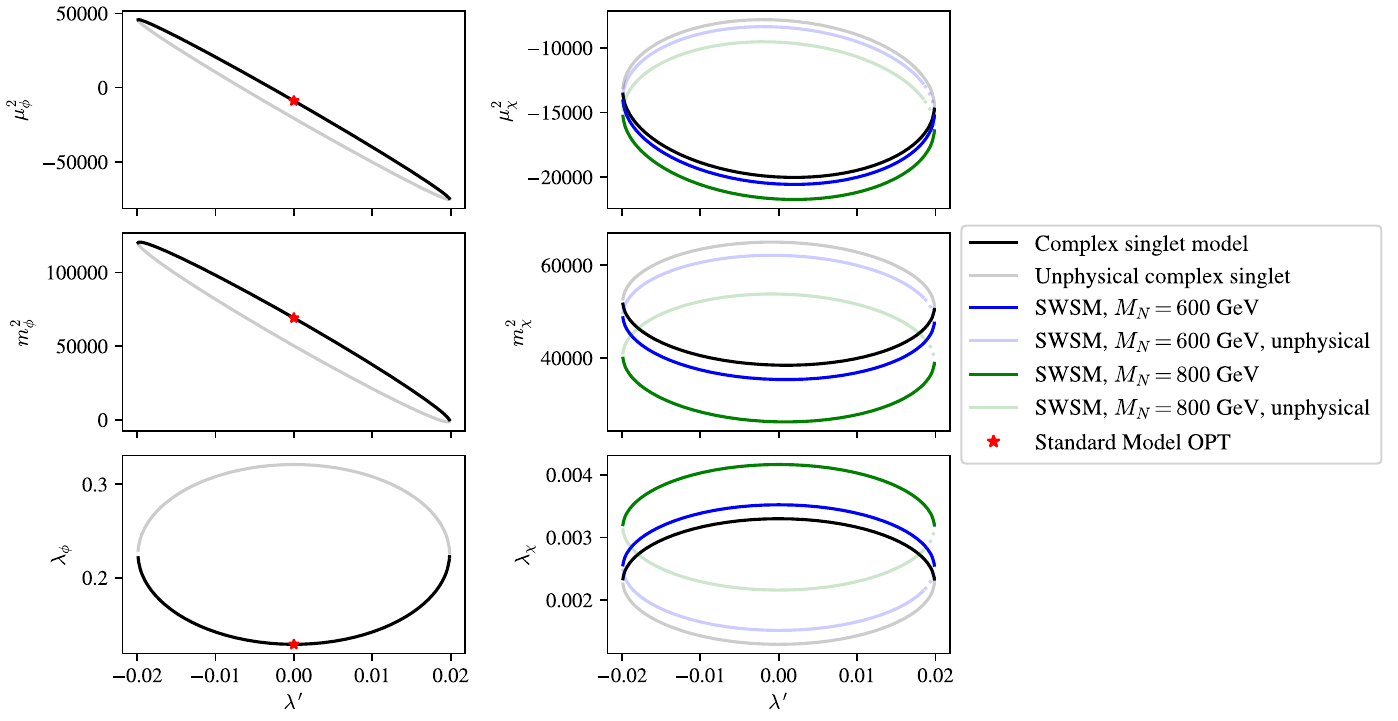}
    \caption{Parametrization of the SWSM in the OPT scheme.
    We used $w_0=10v_0$, $M_s=200~$GeV, and $Q=M_t$ and for simplicity we set $y_{N_2}=y_{N_3}$.
    Only heavy RHNs are shown, as lighter ones do not significantly alter the parameter values.
    On the SM side (left panels) the parameters are unaffected by changing the RHN masses and only the $M_{N}=0$ plot is shown, i.e.,~when the SWSM effective potential reduces to the complex singlet one.}
    \label{fig:SWSMParameters}
\end{figure}

An example parametrization is shown in Fig.~\ref{fig:SWSMParameters} for various RHN masses for a given value of the scalar mass $M_s$ and VEV $w_0$.
The figure shows that the SM sector parameters ($\mu_\phi^2,m_\phi^2$, and $\lambda_\phi$) are indeed independent of the choice for the RHN masses.
The RHNs shift the singlet scalar sector parameters as described by Eq.~\eqref{eq:SWSMparametershift}. 
The SWSM does not change the parametrization significantly as compared to the singlet scalar extension, and we refer to the previous subsection, Sec.~\ref{sec:SingletExtendedSM}, for the qualitative picture of the effects of varying the scalar mass $M_s$ or the VEV $w_0$.

In this section, we have shown the OPT parametrization of the SWSM effective potential.
The tiny gauge coupling renders the description shown here essentially equivalent to any extension which includes extra fermions and a complex singlet scalar.

\section{Effective potential at one loop and finite temperature \label{sec:finT_pot}}

The finite temperature corrections to the effective potential are given by \cite{Dolan:1973qd}
\begin{equation}
    V^{(T)}(T,m_i^2) = \sum_i n_i \frac{T^4}{2\pi^2}J_\pm\left(m_i^2,T\right)\,,
\end{equation}
where $m_i^2$ is the background dependent tree-level mass.
The sum involves in principle all particles in the theory.
Similarly to the zero temperature case, we can safely neglect the contribution of the light modes, and only consider the heavy degrees of freedom.

To cure the IR divergency produced by the static ($n=0$) Matsubara mode, it is costumary to perform a daisy resummation on the bosonic modes, which then adds a thermal Debye mass to their zero temperature one \cite{Carrington:1991hz}.
Note that for the gauge bosons only the longitudinal masses are modified, since at high temperature the leading order of the transverse part of the self-energy is $\propto mT$ whereas for the longitudinal part it is $\propto T^2$ \cite{Espinosa:1992kf}.
Following standard notation, we denote the mass eigenvalues including the thermal contributions (i.e.~the thermal masses) as $\overline{m}_i$.
We give the thermal masses in Appendix~\ref{app:ThermalMasses}.
In the Arnold-Espinosa approach \cite{Arnold:1992rz} the thermal functions $J_\pm(m^2,T)$ are given by
\begin{equation}
    \label{eq:Jplusminus}
    J_\pm\left(m_i^2,T\right) = 
    \begin{cases}
        \displaystyle
        \mathcal{I}_-\left(\frac{m_i^2}{T^2}\right) - \frac{\pi}{6}\left(\frac{\overline{m}_i^3}{T^3}-\frac{m_i^3}{T^3}\right)\,, & \text{if } i=\text{ scalars, longitudinal modes}\,, \\[7pt]
        \displaystyle
        \mathcal{I}_\pm\left(\frac{m_i^2}{T^2}\right)\,,&\text{if }i=\text{ fermions/transverse modes}\,.
    \end{cases}
\end{equation}
For example, in the SM the longitudinal modes would correspond to $W_\rL,Z_\rL,$ and $\gamma_\rL$, the scalars are the Higgs and the Goldstone bosons, while $W_\rT,Z_\rT$ are the transverse modes.
Even though the photon is massless at zero temperature, its longitudinal component (which does not even exist in the vacuum) gets a non-zero contribution at finite temperature.
We introduced $\mathcal{I}_\pm$ as the fermionic ($+$) or bosonic ($-$) integral
\begin{equation}
    \label{eq:Iplusminusintegral}
    \mathcal{I}_\pm\left(a^2\right)=\mp \int_0^\infty\rd x~ x^2\log\left[1\pm e^{-\sqrt{x^2+a^2}}\right]\,.
\end{equation}
The numerical evaluation of this integral for any $T$ and background field value is computationally expensive, so we use the semi-analytic method described in Sec. 2.2 of Ref.~\cite{Basler:2016obg} to approximate its result.
The idea of this approximation is to expand the integral for small and large positive values of $a^2$ \cite{Anderson:1991zb}, then connect the two series continuously at some particular intermediate point $a_0$ where the derivatives match.
Using sufficiently many terms in both expansions, the analytic result matches that of the numerical one to great accuracy.

With increasing temperature the minimum position of the potential
changes, and at sufficiently large temperatures the system eventually settles in its symmetric phase, i.e.~all VEVs become zero.
The order of the phase transition from the broken phase to the symmetric one and the transition temperature $T_{\rm c}$ can be estimated by analyzing the potential at finite temperature.
As described in the introduction, a strongly first order phase transition from tree level effects is excluded in a singlet scalar extension of the SM where the broken phase has a non-zero singlet VEV.
Then only thermal effects create a barrier between the degenerate symmetric- and broken-phase minima, and two competing effects have to be considered for the EWPT. 
On one hand, requiring large couplings of the singlet to the Higgs boson results in heavy ($M_s\gg T_{\rm c}$) singlets that do not have any significant effect at EWPT due to decoupling.
On the other hand, a lighter scalar ($M_s=\mathcal{O}(T_{\rm c}$)) is not decoupled, but consequently it has a small coupling to the Higgs and its contributions are negligible.
A strongly first order EWPT is thus not expected in our investigations of the singlet extension, and we will not consider it.
We emphasize however, that the OPT approach can be employed straightforwardly in models displaying a strong first order EWPT, such as singlet extensions of the SM with vanishing singlet VEV.

In order to be able to track the evolution of the temperature dependent minimum of the effective potential, we need the potential to be real at any value of the background fields.
In the symmetric phase (at high temperature) the vacuum expectation values of the scalar fields vanish and the minimum of the potential is at the origin.
However, in conventional calculations the mass squared parameter is negative and the potential is complex below the zero temperature minimum (see Eq.~\eqref{eq:V_SM_pole-param}), and minimization of the potential is not defined here unless the real part is taken.
Without such a procedure the symmetric phase cannot be recovered.

In the OPT scheme the masses of the scalars and Goldstone bosons could be made real for all values of the background fields at $T=0$.
Consequently the effective potential is generally real and its minimzation is well defined.
In the leading order of the high temperature expansion (HTE) the thermal corrections to particle masses are given in Appendix~\ref{app:ThermalMasses}.
For fields where the mass eigenstates differ from those originally in the Lagrangian (gauge eigenstates), thermal masses need to be calculated by diagonalizing the full temperature-dependent mass matrix.
However, the thermal contributions to the scalar mass squared may be negative if $\lambda'<0$, leading to an imaginary mass (and effective potential) at sufficiently high temperatures.
Imaginary thermal masses are unphysical and the requirements for their reality can be viewed as constraints on the parameter space.
These constraints are tenuous as there is nothing inherently wrong with having negative contributions in the thermal mass \cite{Baldes:2018nel}.
The appearance of the imaginary masses can be interpreted in two ways: either the given parameters are unphysical and should be excluded, or the model in question is not sensible at these temperatures (which sets the energy scale).

\begin{figure}[t]
    \centering
    \includegraphics[width=0.8\linewidth]{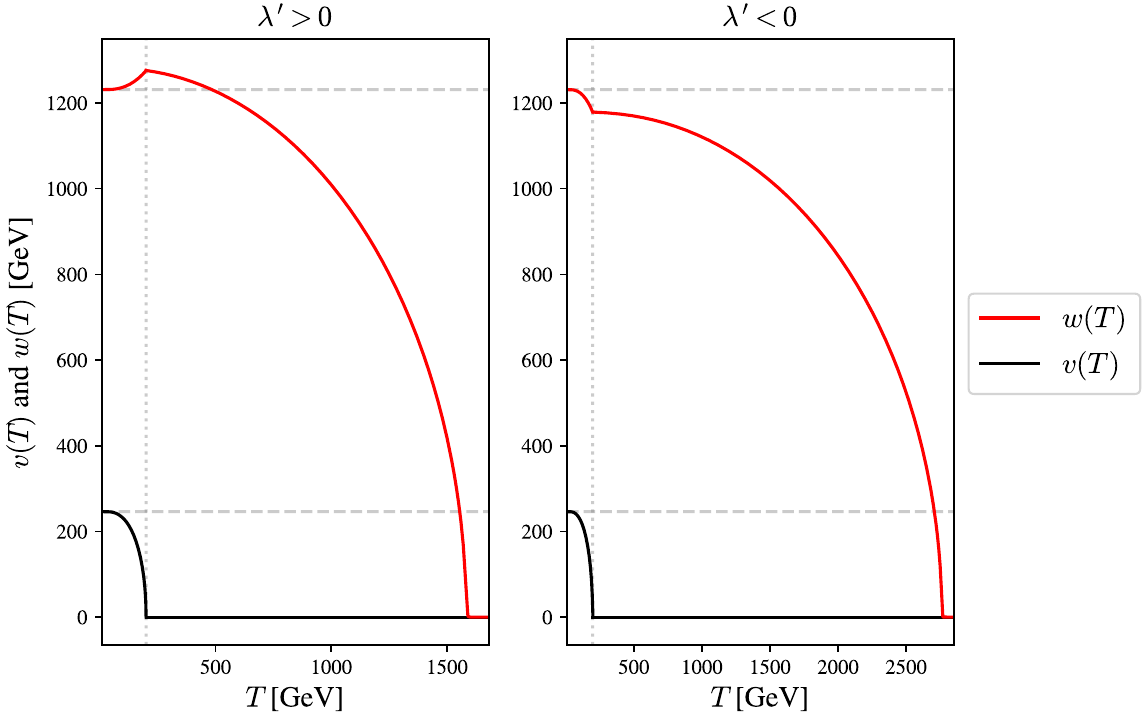}
    \caption{Thermal evolution of the minimum of the effective potential in the SWSM.
    In the figure $M_s=200\GeV$, $M_{N}=150\GeV$, $w_0=5v_0$, and $|\lambda'|=0.0394$.
    With these parameters and within the OPT scheme the effective potential is real.
    The two panels differ in the sign of the mixing: in the left panel $\lambda'>0$ and $w_0$ is approached from above, while in the right panel $\lambda'<0$ and $w_0$ is approached from below as $T\to 0$.
    Note that apart from the qualitative difference at low temperatures, there is also a quantitative difference between the critical temperatures.
    }
    \label{fig:ThermalVEVs}
\end{figure}

We present the thermal evolution of the minimum of the potential in the SWSM in Fig.~\ref{fig:ThermalVEVs}. 
A benchmark point was chosen with $M_s=200\GeV$, $M_N=150\GeV$, and $w_0=5v_0$ where the absolute value of the mixing was fixed to approximately its maximal value, $|\lambda'|=0.0394$.
Mixing indicates that the scalar fields that appear in the Lagrangian are not mass eigenstates, and we have to diagonalize the thermal mass matrix at every temperature to obtain the real masses as eigenvalues.
The sign of the mixing also matters, as the evolution at late times (at low temperatures) is qualitatively different: positive mixing (left panel) implies that $w(T)$ approaches its $T=0$ value from above, while for negative mixing (right panel) $w_0$ is approached from below as $T\to 0$ \cite{Chiang:2017nmu}.

\begin{figure}
    \centering
    \includegraphics[width=0.495\linewidth]{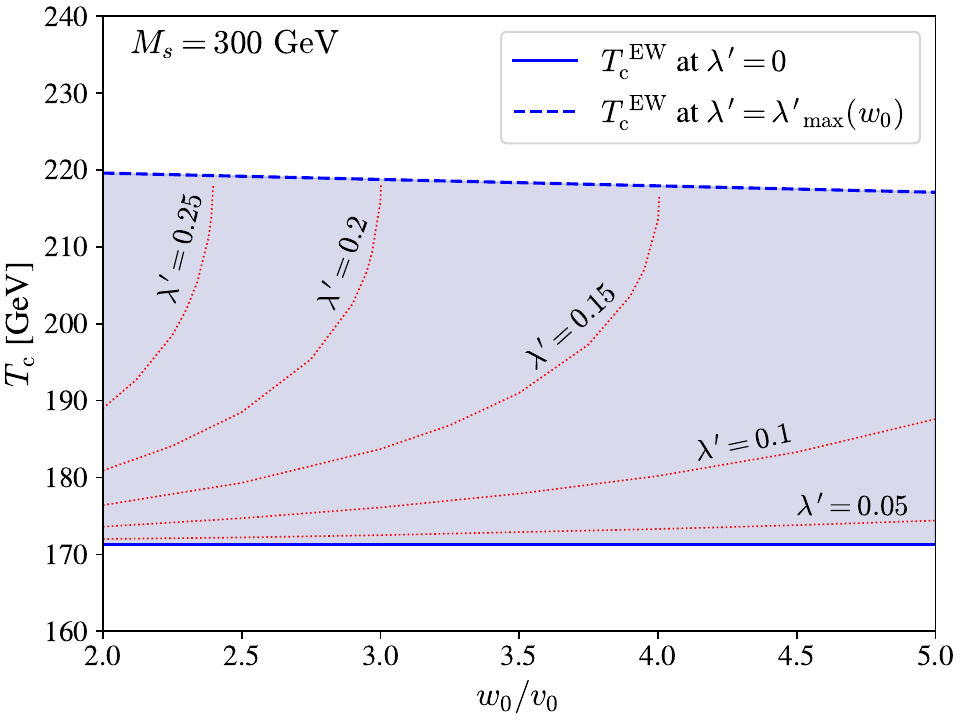}
    \includegraphics[width=0.495\linewidth]{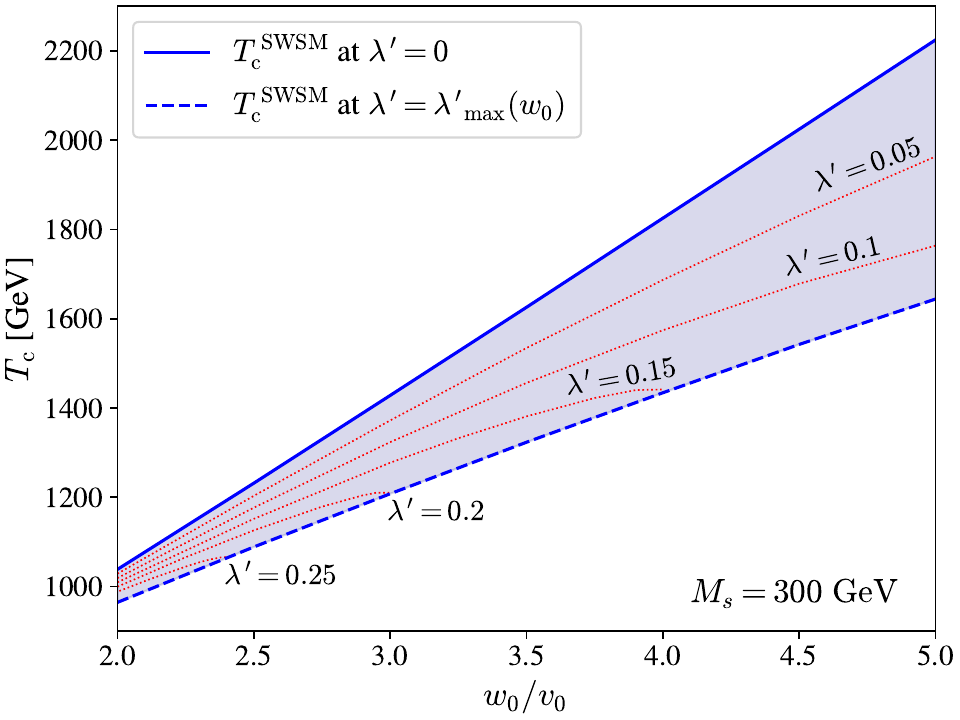}
    \caption{Critical temperatures for the EW ($T_c^{\mathrm{EW}}$, left panel) and SWSM phase transitions ($T_c^{\mathrm{SWSM}}$, right panel) for fixed singlet scalar mass $M_s$, and RHN mass $M_N=150~$GeV.
    The blue shaded region indicates the possible critical temperature ranges.
    The red dotted lines are the critical temperatures at constant values of the scalar mixing parameter $\lambda'$, while the blue dashed line shows the critical temperatures calculated at $\lambda'_\mathrm{max}$, which is a function of $w_0$ (c.f.~Eq.~\eqref{eq:LambdahsMax}).}
    \label{fig:Tc_300_Combined}
\end{figure}

While the overall shape of the thermal evolution of the minimum of the effective potential remains the same as that shown in Fig.~\ref{fig:ThermalVEVs}, the value of the critical temperature for each phase transition depends on the model parameters.
The critical temperatures in the SWSM depend on four parameters: the masses of the singlet scalar and the RHN $M_s$ and $M_N$, the singlet scalar VEV $w_0$, and the scalar mixing $\lambda'$.
The $\lambda'$ parameter takes values in a finite interval given by Eq.~\eqref{eq:LambdahsMax}. Thus by fixing the masses we can study the critical temperatures as functions of only $w_0$ while varying $\lambda'\in[0,\lambda'_{\mathrm{max}}]$ (we focus on positive $\lambda'$ for simplicity to avoid possible imaginary thermal contributions).
We show one such dependence in Fig.~\ref{fig:Tc_300_Combined} for $M_s=Q=300~$GeV and $M_N=150~$GeV.
The phase transitions happen within a temperature band depending on the value of $\lambda'$.
The EWPT critical temperature $T_{\mathrm{c}}^{\mathrm{EW}}$ depends weakly on $w_0$ even for larger values of the scalar sector mixing $\lambda'$.
On the contrary, the SWSM phase transition $T_{\mathrm{c}}^{\mathrm{SWSM}}$ changes roughly linearly with increasing $w_0$.
The observation of phenomenological relevance depicted in the right panel of Fig.~\ref{fig:Tc_300_Combined} is that the SWSM phase transition had to happen at or above temperatures of $\mathcal{O}(1)$ TeV even for a light (but heavier than $M_h$) new scalar.

\section{Conclusions}

In this paper we applied the OPT scheme for the construction of a one-loop effective potential that is real for all values of the background field or fields.
We first showed the viability of this approach on the SM effective potential by comparing it to a benchmark potential, then explored BSM scenarios.
In particular, we studied in detail the parametrization of the SM with an additional singlet scalar, and looked at the finite temperature behavior of the effective potential in the SWSM.
It is straightforward to extend the OPT method for other models with multiple scalar fields.

We constructed the benchmark SM potential in Landau gauge by revisiting the method presented in Ref.~\cite{Boyd:1992xn} to tackle the infrared divergence in the second derivative of the SM one-loop effective potential at zero temperature.
In our approach we not only imposed the value of the pole mass of the one-loop Higgs propagator at the minimum of the potential, but also required that the pole should have a unit residue.
As a result of the finite wave function renormalization, this procedure provides an IR finite one-loop effective potential without residual dependence on the regularization scale of the dimensional regularization scheme used, as opposed to the case when the unity of the residue is not imposed.
The OPT potential reproduces this benchmark potential to high relative accuracy (within 1\,\%) even for temperatures where the EW phase transition takes place. 
We expect that after fixing the gauge, it is safe to use the OPT approach for the calculation of the effective potential in order to estimate the critical temperatures in phase transitions occurring also in scalar extensions of the SM.

We have explored phase transitions with the finite temperature effective potential of the SWSM via the OPT approach.
By restricting our search to two-step phase transitions $(0,0)\to(0,w')\to(v,w)$ with $w\neq0$, we found that for a light (but heavier than $M_h$) new singlet scalar the EWPT depends only slightly on the BSM scalar sector parameters, whereas the singlet phase transition happens at temperatures at or above $T=\mathcal{O}(1)~$TeV.
Moreover, the EWPT is first order (as in the SM) but too weak to provide baryogenesis in any region of the parameter space of the SWSM, as expected from general considerations of singlet extensions.
However, the relatively high scale of the SWSM phase transition may provide ample time for leptogenesis.
This will be investigated in a forthcoming work.

\acknowledgments
We thank G\'abor Cynolter, Gergely Fej\H os, Matteo Giordano, Sho Iwamoto, and Zolt\'an P\'eli for useful discussions.

\appendix 

\section{One-loop Higgs self-energy in the Landau gauge \label{app:SE}}

The Higgs self-energy can be expressed in terms of $d$-dimensional tadpole and bubble integrals. Using the propagator $D(q;m)=\ri/(q^2-m^2 - \ri 0^+)$, these integrals are defined as 
\begin{equation}
  \mathcal{T}(m)=Q^{4-d}\int\frac{\rd^d q}{(2\pi)^d} D(q;m), \quad
  \mathcal{B}(p^2;m,M)=-\ri Q^{4-d}\int\frac{\rd^d q}{(2\pi)^d} D(q;m) D(p-q;M)\,.
\end{equation}

Keeping only those degrees of freedom of the SM that are relevant for the EW phase transition, the self-energy of the Higgs boson gets contributions from the massive gauge bosons, fermions, Goldstone bosons, and through self-interaction:
\begin{subequations}
\begin{equation}
\Pi(p^2;v)=\Pi_\mathrm{tad}(v)+\Pi_\mathrm{bub}(p^2;v)\,,
\end{equation}
\begin{equation}
    \Pi_{\rm tad}(v) = 3\lambda \big[\mathcal{T}(m_h) + \mathcal{T}(m_G)\big]
  -6 y_t^2 \mathcal{T}(m_t) 
  + (d-1)\bigg[\frac{\gL^2}{2}\mathcal{T}(m_W) + \frac{\gZ^2}{4}\mathcal{T}(m_Z)\bigg] \,,  
\end{equation}
\begin{equation}
\begin{split}
  \Pi_{\rm bub}(p^2;v) &=  2 \lambda^2 v^2 \big[9 \mathcal{B}(p^2;m_h) + 3 \mathcal{B}(p^2;m_\chi)\big] - 3y_t^2 (4 m_t^2 - p^2)\mathcal{B}(p^2;m_t) 
  \\  &
  +\bigg\{\gL^2\bigg[ m_W^2 (d-1) \mathcal{B}(p^2;m_W) - \frac{p^2}{2}\Big( 2 \mathcal{B}(p^2;m_W) 
  + \mathcal{B}(0;m_W,0)\Big) 
  \\  & \qquad
  + \frac{p^4}{4m^2_W}\Delta_m\mathcal{B}(p^2;m_W) \bigg] \bigg\}
\quad + \quad \bigg\{\gL^2\to \frac{\gZ^2}{2}\, ,\  m_W\to m_Z \bigg\}\,.
 \label{eq:Pi_bub}
\end{split}
\end{equation}
\end{subequations}
Here we introduced $\mathcal{B}(p^2;m)\equiv \mathcal{B}(p^2;m,m)$ and $\Delta_m \mathcal{B}(p^2;m) \equiv \mathcal{B}(p^2;m)-\mathcal{B}(p^2;0)$ for the finite difference. The $v$-dependence of the tree-level masses was not indicated in order to alleviate the notation. Cancellation occurred between contributions from bubble diagrams involving momentum-dependent vertices and those containing non-derivative interactions.

In the $\overline{\rm MS}$ scheme one obtains the following expansions in $\varepsilon$ for $d=4-2\varepsilon$:
\begin{eqnarray}
\label{eq:Tad_MS_bar}  
16\pi^2\mathcal{T}(m)&\!=\!&m^2 \left(D_{\overline{\rm MS}} -1 +\ln\frac{m^2}{Q^2}\right) +\mathcal{O}(\varepsilon),\\
\label{eq:Bub_MS_bar}
16\pi^2\mathcal{B}(p^2;m, M)&\!=\!& D_{\overline{\rm MS}} +\! \int_0^1 \!\!\rd x
\ln\bigg[(1-x)\frac{m^2}{Q^2}+x \frac{M^2}{Q^2} - \ri 0^+ - \frac{p^2}{Q^2} x(1-x)\bigg] + \mathcal{O}(\varepsilon),\qquad 
\end{eqnarray}
where $D_{\overline{\rm MS}}$, representing the UV divergence, is given below Eq.~\eqref{eq:V_SM_CW}. 

The $p$-dependent divergence of $\Pi_{\rm bub}$ is produced by contributions of the top quark and gauge bosons in which the bubble integral is multiplied by $p^2$. These are removed by
\begin{equation}
\label{eq:dZ_def}
\delta_Z =  \frac{\partial}{\partial p^2}\Pi_{\rm bub}^{\rm div}(p^2;v)=
\frac{D_{\overline{\rm MS}}}{16\pi^2}\,
\bigg(3 y_t^2 - \frac{3}{2} \gL^2 - \frac{3}{4} \gZ^2\bigg) . 
\end{equation}
The $p$-independent divergence of $\Pi_{\rm bub}$ is proportional to $v^2$, hence it can be removed by $2v^2\delta_\lambda$ (see Eq.~\eqref{eq:2pf}), with the coupling counterterm
\begin{equation}
\delta_\lambda = 
-  \frac{D_{\overline{\rm MS}}}{16\pi^2}\,
\bigg(12\lambda^2 - 3 y_t^4 + \frac{3}{8}\gL^4 + \frac{3}{16}\gZ^4\bigg).
\end{equation}
One can check using Eqs.~\eqref{eq:Tad_MS_bar} and \eqref{eq:SMmassesScalars} that $\delta_\lambda$ consistently cancels also the $v$-dependent divergence of $\Pi_{\rm tad}$. Its remaining divergence is removed by the mass counterterm
\begin{equation}
  \delta_{\mu^2}=
  -6\lambda \mu^2\frac{D_{\overline{\rm MS}}}{16\pi^2}\,. 
\end{equation} 

Using the correspondence $\displaystyle D_{\overline{\rm MS}}\rightarrow1+\ln(M^2/\Lambda^2)$ (see e.g.~\cite{Cynolter:2010ei}) where $M$ and $\Lambda$ are the renormalization and regularization (here cutoff) scales, and Eqs.~(12.50) and (12.53) of Ref.~\cite{Peskin:1995ev}, namely 
\begin{equation}
\gamma^{(1)}=\frac{1}{2}M\frac{\partial \delta_Z}{\partial M},\quad
\beta^{(1)}_\lambda=M\frac{\partial}{\partial M}\bigg(-\delta_\lambda+4\frac{\lambda}{2}\delta_Z\bigg),\quad
\beta^{(1)}_{\mu^2}=M\frac{\partial}{\partial M}\bigg(-\delta_{\mu^2}+2\frac{\mu^2}{2}\delta_Z\bigg),
 \end{equation} 
one can verify that the above counterterms give the correct expression of the one-loop $\gamma$ and $\beta$ functions, appearing for instance in Eqs.~(A34), (A35), and (A36) of Ref.~\cite{Martin:2014bca}.

\section{Construction of the SM potential based on the Higgs pole mass \label{app:PoleBasedPot}}

In this Appendix we work out the procedure outlined in Sec.~\ref{sec:PoleMassparametrization}, giving the ingredients needed to reproduce the effective potential \eqref{eq:V_SM_pole-param}.

With the notation introduced in Eq.~\eqref{eq:L_CT_notation}, the one- and two-point functions of the Higgs field have the following form at one-loop level 
\begin{eqnarray}
  \label{eq:1pf}
  0 &=& v \Big[\mu_R^2 + \lambda_R v^2 + \Pi_\mathrm{tad}(v) + \delta_{\mu^2} + v^2\delta_\lambda\Big]
  \,,\\
  \label{eq:2pf}
\ri G_h^{-1}(p^2;v) &=& p^2 - 2\lambda v^2 - \Big[\mu_R^2 + \lambda_R v^2 + \Pi_\mathrm{tad}(v) + \delta_{\mu^2} + v^2\delta_\lambda\Big] \nonumber\\
&&- \Big[2 v^2\Delta\lambda -p^2 \Delta Z + \Pi_\mathrm{bub}(p^2;v) + 2 v^2\delta_\lambda - p^2\delta_Z\Big]
\,,
\end{eqnarray}
where we used that the one-loop self-energy is a sum of tadpole and bubble diagrams. We also used that the contribution of the 1PI tadpole diagrams appearing in the two-point function multiplied by $v$ is the corresponding tadpole contribution to the one-point function.
The infinite counterterms are used such that each individual square bracket is finite, which can be explicitly checked using the formulae of Appendix~\ref{app:SE}. The finite part of any contribution to the self-energy is denoted in what follows by the superscript F, e.g.,
\begin{equation}
    \Pi_\mathrm{bub}(p^2;v)  = \Pi^\mathrm{div}_\mathrm{bub}(p^2;v) + \Pi^\mathrm{F}_\mathrm{bub}(p^2;v) 
    \,.
\label{eq:div+F}
\end{equation}
Such a decomposition is scheme dependent and we use the $\overline{\rm MS}$ subtraction scheme, as in Appendix~\ref{app:SE}.

The requirement that the minimum of the one-loop potential remains where it was in the tree-level one is equivalent to the requirement that $v_0$, satisfying $\mu^2+\lambda v_0^2=0$, is the nontrivial solution of \eqref{eq:1pf}, which implies that
\begin{equation}
  \label{eq:rel_Dm_Dl}
  \Delta \mu^2 + v_0^2 \Delta \lambda + \Pi^{\rm F}_\mathrm{tad}(v_0) = 0
  \,.
\end{equation}
Two more relations between the finite parameters can be obtained from the Higgs propagator evaluated at the minimum $v_0$ of the potential. We use Taylor expansion
\begin{equation}
  \Pi^{\rm F}_\mathrm{bub}(p^2;v_0)=\Pi^{\rm F}_\mathrm{bub}(M_h^2;v_0) + (p^2-M_h^2)  \Pi'^{\rm F}_\mathrm{bub}(M_h^2;v_0) + \tilde \Pi^{\rm F}_\mathrm{bub}(p^2,M_h^2;v_0)
\,,
\end{equation}
with $\tilde \Pi^{\rm F}_\mathrm{bub}$ defined by the equality, and exploit the tree-level relation $M_h^2=2\lambda v_0^2$ to obtain
\begin{eqnarray}
  \ri G_h^{-1}(p^2;v_0)&=&[p^2 - M_h^2] \Big[1-\Big(\Pi'^{\rm F}_\mathrm{bub}(M_h^2;v_0) - \Delta Z\Big)\Big]\nonumber \\
  && - \Big(2\Delta\lambda v_0^2 + \Pi^{\rm F}_\mathrm{bub}(M_h^2;v_0) - M_h^2 \Delta Z\Big) - \tilde \Pi^{\rm F}_\mathrm{bub}(p^2,M_h^2;v_0).
\end{eqnarray}
Then, the requirement that the expressions in the two parentheses vanish, that is
\begin{equation}
  \label{eq:DZ_and_Dl}
  \Delta Z =
  \frac{\partial}{\partial p^2}\Big[\Pi_{\rm bub}(p^2;v_0) - p^2 \delta_Z \Big]\Big|_{p^2=M_h^2} \quad \text{and} \quad
  \Delta\lambda = -\frac{1}{2v_0^2} \Pi^{\rm F}_\mathrm{bub}(M_h^2;v_0) + \lambda \Delta Z, 
\end{equation}
guarantees that $p^2=M^2_h$ is a pole of the Higgs propagator with unit residue due to the limit $\displaystyle\lim_{p^2\to M_h^2}\tilde \Pi(p^2,M_h;v_0)/(p^2-M_h^2) = 0$. In writing $\Delta Z$ we expressed $\Pi'^{\rm F}_\mathrm{bub}(p^2;v_0)$ from Eq.~\eqref{eq:div+F} and used the definition of $\delta_Z$ in Eq.~\eqref{eq:dZ_def}.  
Conditions similar to those in \eqref{eq:DZ_and_Dl} were used in Ref.~\cite{Appelquist:1973ms} for the renormalization of the Abelian-Higgs model. We note that while $\delta_Z$ receives contribution only from the top quark and gauge bosons, all modes contribute to the finite wave function renormalization factor $\Delta Z$.

The renormalized one-loop potential can be constructed using $\Delta\mu^2$ and $\Delta\lambda$, given in Eqs.~\eqref{eq:rel_Dm_Dl} and \eqref{eq:DZ_and_Dl}, and the infinite counterterms $\delta_{\mu^2}$ and $\delta_\lambda$, but bringing the result in the form presented in Eqs.~\eqref{eq:V_SM_pole-param} proves difficult.
A better strategy is to work with bare couplings. The classical potential written in terms of bare couplings reads
\begin{equation}
    V_{\rm cl,0}(v;\mu_0^2)=\Omega_0+\frac{\mu^2_0}{2}v^2 + \frac{\lambda_0}{4}v^4\,,
\end{equation}
where we separated a (bare) constant $\Omega_0$ for convenience.
We fix $\Omega_0$ from the condition $V^{[1]}(v_0)=0$, which results in
\begin{equation}
\label{eq:Vtree_bare}
V_{\rm cl,0}(v;\mu_0^2)=\frac{\mu^2_0}{2}(v^2-v_0^2) + \frac{\lambda_0}{4}(v^4-v_0^4) - V^{(1)}(v_0;\mu^2).
\end{equation}

Our goal is to express the bare couplings $\mu_0^2$ and $\lambda_0$ and bring the sum of \eqref{eq:Vtree_bare} and \eqref{eq:V_SM_1loop} into the form \eqref{eq:V_SM_pole-param}. The equations determining $\mu_0^2$ and $\lambda_0$ are
\begin{subequations}
  \label{eq:eqs_for_m0_l0}
  \begin{eqnarray}
    \mu_0^2 + \lambda_0 v_0^2 + \Pi^G_{\rm tad}(v_0) + \frac{1}{v_0}\frac{\rd}{\rd v_0}\bar V^{(1)}(v_0;\mu^2) &=& 0,
    \\
    \label{eq:eqs_for_m0_l0_from_2pf}
    \mu_0^2 + 3\lambda_0 v_0^2 + \Pi(M_h^2;v_0)-M_h^2\frac{\partial}{\partial p^2}\Pi_{\rm bub}(p^2;v_0)\Big|_{p^2=M_h^2} &=& M_h^2.
  \end{eqnarray}
\end{subequations}
The first equation is \eqref{eq:1pf}, written in terms of bare couplings. We separated in it the contribution of the Goldstone bosons. The contribution of all the other modes is written using the comment below \eqref{eq:2pf} and is denoted with a {\it bar} on the quantity they contribute to, e.g. $\displaystyle \bar V^{(1)}(v,\mu^2) = \sum_{i\ne G} n_i V^{(1)}_{\rm CW}(v,m_i^2(v,\mu^2))$. The second equation is obtained from $\ri G_h^{-1}(M_h^2;v_0) = 0$, using the first equation of \eqref{eq:DZ_and_Dl} and the relations in (and below) Eq.~\eqref{eq:cpl_decomp} on the right hand side of Eq.~\eqref{eq:2pf}. 

In order to relate \eqref{eq:eqs_for_m0_l0_from_2pf} to the effective potential without introducing IR divergences, one should separate the contribution of the Goldstone modes. Using that the one-loop contribution of each non-Goldstone mode to the Higgs curvature mass squared is equal to its contribution to the one-loop self-energy evaluated at vanishing external momentum, one has the identity
\begin{equation}
  \label{eq:IR-safe_identity}
  \Pi(M_h^2;v_0) = \Pi^G(M_h^2;v_0) + \frac{\rd^2}{\rd v_0^2} \bar V^{(1)}(v_0;\mu^2) + \big[\bar\Pi(M_h^2;v_0) - \bar\Pi(0;v_0)\big],
\end{equation}
in which all terms are IR finite. Note that the contribution of the tadpoles cancels in the square bracket: $\bar\Pi(M_h^2;v_0) - \bar\Pi(0;v_0) =  \bar\Pi_{\rm bub}(M_h^2;v_0) - \bar\Pi_{\rm bub}(0;v_0)$. We denote this finite difference by $\Delta\bar\Pi_{\rm bub}(M_h^2;v_0)$.

Using the relation $M_h^2=2\lambda v_0^2$ and Eq.~\eqref{eq:IR-safe_identity}, one can solve the equations in \eqref{eq:eqs_for_m0_l0} for the bare couplings and find
\begin{subequations}
  \label{eq:m0_and_l0}
\begin{eqnarray}
\mu_0^2 &=&\mu^2 - \Pi^G_{\rm tad}(v_0) + \frac{1}{2} \left[\frac{\rd^2}{\rd v_0^2} -\frac{3}{v_0}\frac{\rd}{\rd v_0}\right] \bar V^{(1)}(v_0;\mu^2)-\frac{1}{2}C\,,
\\
\lambda_0 &=& \lambda - \frac{1}{2 v_0^2} \left[\frac{\rd^2}{\rd v_0^2} -\frac{1}{v_0}\frac{\rd}{\rd v_0}\right] \bar V^{(1)}(v_0;\mu^2) - \frac{C}{2 v_0^2}\,,
\end{eqnarray}
where $\Pi^G_{\rm tad}(v_0)=0$ in DR and $C\equiv C(M_h^2;v_0)$ is a particular value of the function
\begin{equation}
\label{eq:C_in_m0_l0}
C(p^2;v_0)=
\bigg(1-M_h^2\frac{\partial}{\partial p^2}\bigg)\Pi^G_{\rm bub}(p^2;v_0) + \Delta\bar\Pi_{\rm bub}(M_h^2;v_0) - M_h^2\frac{\partial}{\partial p^2}\bar\Pi_{\rm bub}(p^2;v_0)
\,.
\end{equation}
\end{subequations}

Having determined the bare couplings, one can use the explicit expression of the Higgs self-energy given in Appendix~\ref{app:SE} to perform the final step of the calculation. 
Adding \eqref{eq:Vtree_bare} to \eqref{eq:V_SM_1loop} and using the contribution of each mode to $\mu_0^2$ and $\lambda_0$ given in \eqref{eq:m0_and_l0}, a straightforward calculation leads to \eqref{eq:V_SM_pole-param}, showing that all UV divergences and dependencies on the regularization scale cancel. 
In the absence of a finite wave-function renormalization a dependence on the regularization scale $Q$ would remain from the contributions of the top quark and gauge bosons to $\Delta\bar\Pi_{\rm bub}(M_h^2;v_0)$. The derivatives $\bar V'^{(1)}$ and $\bar V''^{(1)}$ appearing in the expression of $\mu_0^2$ and $\lambda_0$ are to be rewritten as mass derivatives for each mode involved. They combine with $\bar V^{(1)}(v;\mu^2)-\bar V^{(1)}(v_0;\mu^2)$, giving the expression
appearing in the last term of \eqref{eq:V_SM_pole-param}. This finite contribution of a non-Goldstone mode corresponds to the subtraction of the first three terms of the Taylor expansion of
$\ln(p^2+m_i^2(v))$ about $\ln(p^2+m_i^2(v_0))$ in the defining integral of $V_{\mathrm{CW}}(m_i^2(v))$.

The functions listed in \eqref{eq:l_funcs} 
are obtained from the term proportional to $C$ in \eqref{eq:m0_and_l0}. Considering only the divergent part of the self-energy when taking the derivative in \eqref{eq:C_in_m0_l0} corresponds to $\Delta Z=0$, which is the case considered in Ref.~\cite{Boyd:1992xn}. A detailed comparison shows that for the top quark and the gauge bosons the result given there in (A1) is not in line with the Higgs self-energy computed e.g. in Ref.~\cite{Casas:1994us}, with which we agree. Namely, the last term in the contribution of the top quark appearing there should be $4r$, while concerning the real part of the gauge boson's contribution only the first term seems correct.\footnote{We obtain
$\frac{3}{4}{\rm Re}f_W(r)=(12-4r+r^2)F(r)\arctan{(1/F(r))} -12 +5r -3r\ln\frac{M_h^2}{rQ^2}-\frac{r^2}{2}\ln r$, where $r=M_h^2/m_W^2(v_0)$ and $F(r)=\sqrt{-1+4/r}$.}

A final remark concerns the imaginary part of the self-energy that appears when contribution of individual modes is considered. 
As shown in \cite{Casas:1994us}, the imaginary parts produced by bubble integrals with vanishing mass add up to zero in the self-energy evaluated for $v_0$ and $p^2=M_h^2$. 
In other words, the coefficient of $\mathcal{B}(M_h;0)$ coming both from the Goldstone boson contribution and the last term in the contribution of gauge bosons in \eqref{eq:Pi_bub} vanishes:
\begin{equation}
    2n_G\lambda^2 v_0^2  - \gL^2\frac{M_h^4}{4 m_W^2(v_0)} - \gZ^2\frac{M_h^4}{8 m_Z^2(v_0)}=\frac{M_h^4}{v_0^2}\bigg[\frac{n_G}{2} - 1 -\frac{1}{2}\bigg] = 0.
\end{equation}
Other bubble integrals do not produce an imaginary part when evaluated with physical masses at $v_0$ and $p^2=M_h^2$. Therefore, only the real part of the bubble integral has to be considered when working with the contribution of individual modes to the self-energy.

\section{Thermal masses}
\label{app:ThermalMasses}

In this appendix we give the thermal masses of various types of fields, calculated at leading order in HTE.
First, general formulae are given for the different contributions, then these formulae are applied to the SM fields.
Similar discussions can be found in Refs.~\cite{Katz:2014bha, Comelli:1996vm}.
For relations regarding the SU($N$) algebra we take the notation and formulae presented in Ref.~\cite{Haber:2019sgz}.

We take the following general Lagrangian for a fermion $\psi$ that transforms under some local gauge group:
\begin{equation}
    \label{eq:GeneralLagrangian}
    \mathcal{L} = \Bar{\psi}_i\gamma^\mu(\ri\partial_\mu \delta_{ij}-g\mathbf{T}^a_{ij}A^a_\mu)\psi_j - \frac{1}{4}F_{\mu\nu}^a F^{\mu\nu}_a - y_\psi\Bar{\psi}\phi\psi\,,
\end{equation}
where $\mathbf{T}^a$ are the generators of the group in the fundamental representation and $g$ is the corresponding generic coupling between the fermions and the gauge fields.

The SM is a gauge theory with SU(3)$_\mathrm{c}\otimes$SU(2)$_\rL\otimes$U(1)$_Y$ local symmetries where the gauge covariant derivative of a field $f(x)$ is
\begin{equation}
\label{eq:CovariantDerivativeGeneralSM}
    \mathcal{D}_\mu f_{ia} = \Big[\partial_\mu\delta_{ij}\delta_{ab} - \ri \gY \delta_{ij}\delta_{ab} Y_f B_\mu - \frac{\ri \gL}{2} (\Vec{\sigma} \cdot \Vec{W}_\mu)_{ij}\delta_{ab} - \frac{\ri \gs}{2} \delta_{ij}(\Vec{\lambda} \cdot \Vec{A}_\mu)_{ab}\Big] f_{jb}\,,
\end{equation}
where $\Vec{\sigma}$ denotes the vector of Pauli matrices, $\Vec{\lambda}$ is the vector of the 8 Gell-Mann matrix, $i,j=1,2,3$, and $a,b=1,2,\dots,8$.

The heaviest particle in the SM is the top quark, for which the Yukawa term is
\begin{equation}
    \label{eq:YukawaLagrangianfull}
    \mathcal{L}_Y^{(t)}=-\frac{y_t}{\sqrt{2}}\Bar{t}\phi_1t + \ri\frac{y_t}{\sqrt{2}}\Bar{t}\gamma_5\phi_2 t - \frac{y_t}{2\sqrt{2}}\Big[\Bar{b}(1+\gamma_5)(-\phi_3+\ri\phi_4)t-\Bar{t}(1-\gamma_5)(\phi_3+\ri\phi_4)b\Big]\,.
\end{equation}
In the terms involving the fields $\phi_{3,4}$ the bottom quark field $b$ appears (we neglect mixing present in the $d$-type quarks as we focus only on the top-bottom quark doublet).
The mass of the top quark $m_t=y_t v/\sqrt{2}$ is generated by the Higgs mechanism when $\phi_1$ acquires a non-zero vacuum expectation value $v$ (c.f.~first term in the Yukawa Lagrangian).

The scalar potential of the BEH field $\phi$ (parametrized as in Eq.~\eqref{eq:BEHfield}) in the SM involves a quartic term in the scalar fields
\begin{equation}
    \label{eq:SMpotExpanded}
    V_\mathrm{SM}(\phi)\supset\lambda |\phi^\dagger\phi|^2 = \sum_i\frac{\lambda}{4}\phi_i^4 + \sum_{i\neq j}\frac{\lambda}{2}\phi_i^2\phi_j^2\,.
\end{equation}

\subsection{Scalar thermal mass}

The thermal mass of scalar fields is obtained from the self-energy diagrams with fermion, scalar, and gauge boson loops \cite{Carrington:1991hz}.
For a real scalar field with a Yukawa coupling to a fermion such as that in Eq.~\eqref{eq:GeneralLagrangian}, the thermal contribution is
\begin{equation}
    \Delta m^{2\,(\psi)}_\phi=\frac{y_\psi^2 T^2}{6}\,.
\end{equation}
In the SM, the Yukawa Lagrangian for the top quark is given in Eq.~\eqref{eq:YukawaLagrangianfull}.
It is seen, that $\phi_{1,2}$ and $\phi_{3,4}$ have different vertices, nevertheless their thermal contribution is the same 
\begin{equation}
    \Delta m^{2\,(t)}_{\phi_i}=\frac{y_t^2 T^2}{4}\,,\quad i=1,2,3,4.
\end{equation}

Given a scalar potential for a real field $\phi$ of the form $V(\phi)=\lambda\phi^4/4$, the two point function gets contributions from scalar self-interactions and yields the thermal contribution to the mass:
\begin{equation}
    \Delta m^{2\,(\phi)}_\phi = \frac{\lambda T^2}{4}\,.
\end{equation}
In the SM the potential is given in Eq.~\eqref{eq:SMpotExpanded}.
The self-energy for any $\phi_i$ ($i=1,\,2,\,3,\,4$) is a sum of a tadpole involving the same field $\phi_i$ and the 3 tadpoles involving $\phi_{i\neq j}$ in the loop.
The thermal contribution is 
\begin{equation}
    \Delta m_{\phi_i}^{2\,(\phi)}=\frac{\lambda T^2}{4} + 3\cdot \frac{\lambda T^2}{12} = \frac{\lambda T^2}{2}\,.
\end{equation}

The gauge boson contributions to the self-energy are due to the kinetic term of the bosons (c.f.~Eq.~\eqref{eq:CovariantDerivativeGeneralSM}).
The couplings between the scalar field and the gauge boson is proportional to the gauge boson mass $M$.
Let $M^2(\phi)=g^2\phi^2/4$, then 
\begin{equation}
    \Delta m_{\phi}^{2\,(\mathrm{gb.})}(M^2)= \frac{M^2(v)T^2}{4v^2}\,.
\end{equation}
In the SM, only the $W^\pm$ and $Z^0$ gauge bosons are massive, thus 
\begin{equation}
    \Delta m_{\phi}^{2\,(\mathrm{gb.})} = 2\Delta m_\phi^{2\,(\mathrm{gb.})}(m_W^2) + \Delta m_\phi^{2\,(\mathrm{gb.})}(m_Z^2) =  \frac{\gL^2T^2}{8} + \frac{\gZ^2T^2}{16}\,.
\end{equation}
In total, the SM Higgs gets the following thermal contribution to its mass:
\begin{equation}
    \Delta m_h^2 = \left(\frac{y_t^2}{4} + \frac{\lambda}{2} + \frac{\gL^2}{8} + \frac{\gZ^2}{16}\right) T^2\,.
\end{equation}

In the SWSM there exists a mixing term between the scalar fields $\mathcal{L}_{\mathrm{mix}}=\lambda' |\phi|^2|\chi|^2$ which induces new diagrams to the self-energy.
These contributions are calculated the same way as the ones in the SM where we put a $\phi_{i\neq j}$ into the tadpole of the one-loop $\phi_i$ propagator.
The BEH field is a doublet, it involves 4 real scalar fields, whereas the complex singlet has 2 real scalars, thus the thermal contributions to the self-energy are
\begin{align}
    \label{eq:scalarmixingThermalMass}
    \Pi_\phi^{\rm (mix)} = 2\cdot \frac{\lambda' T^2}{24} = \frac{\lambda' T^2}{12}\qquad \text{and}\qquad
    \Pi_\chi^{\rm (mix)} = 4\cdot \frac{\lambda' T^2}{24} = \frac{\lambda' T^2}{6}\,.
\end{align}
In order to calculate the thermal masses of the physical scalar states, we need to add the one-loop self-energy to the mass matrix and find the eigenvalues.
We mention that the thermal contribution of the $Z'$ gauge boson is negligible due to $M_{Z'}\ll M_Z$, or put differently due to the feeble coupling $\mathcal{O}(g_z^2/\gZ^2)\ll 1$. 

\subsection{Gauge boson thermal mass}

The gauge boson thermal masses are calculated in e.g.\,Refs.~\cite{Carrington:1991hz,Espinosa:1992kf,Buchmuller:1993bq}.
The gauge boson self-energy gets contributions from self-interactions (non-Abelian gauge fields only), fermion loops, and scalar loops.
In the following we present the Debye masses (corresponding to taking the limit $|\mathbf{q}|\to 0$ at $q_0=0$) of the longitudinal modes of gauge fields \cite{Bellac:2011kqa}, obtained at leading order in the HTE.

The contribution due to self-interaction (derivative cubic and quartic vertices) and ghosts is 
\begin{equation}
    \Pi_{00}^{\rm (gb.)}(g) = \frac{\CA g^2 T^2}{3}\,,
\end{equation}
where $\CA$ is the quadratic Casimir of the given SU($N$) group for the adjoint representation, and $g$ represents either $\gL$ or $\gs$ for the SU(2)$_\rL$ and SU(3)$_\mathrm{c}$ groups respectively. For a group with $N\geq 2$ the Casimir is $\CA=2 \TF N$, with $\TF$ being the normalization of the generators in the fundamental representation, for which we use the conventional value $\TF=1/2$ for any $N$.

For the fermionic contribution we have to sum over all fermion states that can enter the loop.
We will relate the self-energy contributions of fermion fields to the well-known QED result
\begin{equation}
    \label{eq:QEDresult}
    \Pi_{00}^{\mathrm{QED}}=\frac{e^2T^2}{3}\,.
\end{equation}
Since QED is a non-chiral theory, the above self-energy is twice the corresponding contribution of an explicit chirality field.
The interactions between gauge fields and fermions can be read from the covariant derivative in Eq.~\eqref{eq:CovariantDerivativeGeneralSM}.

The only fermions that transform non-trivially under the SU(3)$_\mathrm{c}$ group are the quarks.
Given $N_\mathrm{G}=3$ fermion generations (quarks and leptons) there are $N_\mathrm{f}=2N_\mathrm{G}=6$ quark and lepton flavours.
The quark contribution to the gluon self-energy is
\begin{equation}
    \Pi_{00,ab}^{\mathrm{SU(3)}}=N_\mathrm{f}~\mathrm{tr}(\mathbf{t}_a\mathbf{t}_b)\Pi_{00}^{\mathrm{QED}}(e^2\to g_s^2) = \delta_{ab}g_s^2 T^2\,,
\end{equation}
where $\mathbf{t}_a=\lambda_a/2$ ($a=1,2,...,8$) are the half Gell-Mann matrices, and $\mathrm{tr}(\mathbf{t}_a\mathbf{t}_b)=\frac{1}{2}\delta_{ab}$.

To the SU(2)$_\rL$ gauge boson self-energies only the left-handed fermions contribute.
Denoting the number of colors by $N_c$, the SU(2)$_\rL$ self-energy is
\begin{equation}
    \label{eq:SU2selfenergy}
    \Pi_{00,ij}^{\mathrm{SU(2)}}=\frac{1}{2}(N_\mathrm{c}N_{\mathrm{G}}+N_{\mathrm{G}})\mathrm{tr}(\mathbf{T}_i\mathbf{T}_j)\Pi_{00}^{\mathrm{QED}}(e^2\to \gL^2) = \delta_{ij}\gL^2T^2\,,
\end{equation}
where $\mathbf{T}_i=\sigma_i/2$ ($i=1,2,3$) are the half Pauli matrices, and $\mathrm{tr}(\mathbf{T}_i\mathbf{T}_j)=\delta_{ij}/2$.
The factor of $1/2$ in Eq.~\eqref{eq:SU2selfenergy} appears because only
left-handed fermions contribute.

To the U(1)$_Y$ gauge boson self-energy all fermions with non-zero hypercharge contribute.
We denote the hypercharges of any field $i$ by $Y_i$.
Left- and right-handed fields differ under the U(1)$_Y$ group, thus we sum over the specific modes (each separate chiral mode will have a contribution half that of Eq.~\eqref{eq:QEDresult}) of all fermions and find
\begin{align}
    \Pi_{00}^{\mathrm{U(1)}} &= \frac{1}{2}\sum_{i}N_i Y_i^2\Pi_{00}^{\mathrm{QED}}(e^2\to \gY^2) \nonumber\\  
    &=\frac{1}{2}\left[N_\mathrm{c}N_\mathrm{f}Y_Q^2 + N_\mathrm{f}Y_L^2+N_\mathrm{c}N_{\mathrm{G}}(Y_u^2+Y_d^2) + N_{\mathrm{G}}Y_e^2\right]\Pi_{00}^{\mathrm{QED}}(e^2\to \gY^2)  \nonumber\\ 
    &= \frac{5\gY^2T^2}{3}\,.
\end{align}
Here $Y_Q$ ($Y_L)$ is the hypercharge of the left-handed quark (lepton) doublets, while $Y_e,Y_u,Y_d$ denote the hypercharges of the respective right-handed fields.

The scalar contributions only appear for the SU(2)$_\rL$ and U(1)$_Y$ gauge fields, as the BEH field is a color singlet.
The self-energy due to scalars is
\begin{equation}
    \Pi^{(\phi)}_{00}(g)=\frac{g^2T^2}{6}\,,
\end{equation}
where $g$ indicates the coupling corresponding to the given gauge field.

In summary the gauge boson self-energies are as follows:
\begin{subequations}
    \begin{align}
        \Pi^{\mathrm{U(1)}} &= \Pi_{00}^{\mathrm{U(1)}} + \Pi_{00}^{(\phi)}(\gY) = \frac{11}{6}\gY^2T^2\,, \\
        \Pi^{\mathrm{SU(2)}} &= \Pi^{\rm (gb.)}_{00}(\gL) + \Pi^{\mathrm{SU(2)}}_{00,ii} + \Pi_{00}^{(\phi)}(\gL)= \frac{11}{6}\gL^2T^2\,, \\
        \Pi^{\mathrm{SU(3)}} &= \Pi_{00}^{\rm (gb.)}(g_s) + \Pi^{\mathrm{SU(3)}}_{00,aa} = 2g_s^2T^2\,.
    \end{align}
\end{subequations}
In order to obtain the full thermal Debye masses for the longitudinal mode of the gauge bosons, we have to diagonalize their mass matrix with the temperature dependent self-energies included:
\begin{subequations}
    \begin{align}
        \overline{m}_{W_\rL}^2&=\frac{\gL^2v^2}{4}+\frac{11\gL^2T^2}{6}\,, \\
        \overline{m}_{Z_\rL}^2&=\frac{1}{24}\left[\gZ^2(3v^2+22T^2) + \sqrt{9\gZ^4v^4 + 44T^2(\gL^2-\gY^2)^2(3v^2+11T^2)}\right]\,, \\
        \overline{m}_{\gamma_\rL}^2&=\frac{1}{24}\left[\gZ^2(3v^2+22T^2) - \sqrt{9\gZ^4v^4 + 44T^2(\gL^2-\gY^2)^2(3v^2+11T^2)}\right]\,.
    \end{align}
\end{subequations}
In the SWSM, the U(1)$_z$ gauge boson $B'$ weakly mixes with $W_3$ and $B$. 
This mixing modifies the SM thermal masses for $Z_\rL$ and $\gamma_\rL$ only at the level of $\mathcal{O}(g_z^2/\gZ^2)\ll 1$.
Furthermore, the new mass eigenstate $Z'$ receives a thermal mass correction of order $g_z^2$, and thus its contribution to the effective potential is negligible.

\subsection{Fermion thermal mass}

For completeness we also present the fermion thermal masses. For chirally invariant gauge theories they were calculated at high temperature in Ref.~\cite{Weldon:1982bn}.

For a general SU($N$) gauge group with a coupling $g$ to the fermions (c.f.~Eq.~\eqref{eq:GeneralLagrangian}), the contribution of gauge fields to the thermal mass is
\begin{equation}
    \Delta m_\psi^{2\, \mathrm{(gb.)}}=\frac{g^2C_{\rm R}T^2}{8}\,,
\end{equation}
where $C_\mathrm{R}$ is the eigenvalue of the quadratic Casimir operator in the representation R of the fermion field.
In the SM, all fermion fields transform in the fundamental (F) representation.
For a U(1) group $\CF=1$, and for SU($N$) it is given by $\CF=(N^2-1)/2N$.
Note that left- and right-handed fields couple differently to gauge fields, hence their respective thermal masses will differ.

The scalar contribution from a single real scalar field $\phi$ with a Yukawa interaction given in Eq.~\eqref{eq:GeneralLagrangian} is
\begin{equation}
    \Delta m_\psi^{2\,\mathrm{(\phi)}}=\frac{y_\psi^2 T^2}{16}\,.
\end{equation}
The left- (L) and right-handed (R) top quark thermal masses in the SM are
\begin{subequations}
    \begin{align}
        &\Delta m_{t_\rL}^2 = \frac{\gY^2T^2}{288} + \frac{3\gL^2T^2}{32} + \frac{g_s^2T^2}{6} + \frac{y_t^2T^2}{16}\,, \\
        &\Delta m_{t_\rR}^2 = \frac{\gY^2T^2}{18} + \frac{g_s^2T^2}{6} + \frac{y_t^2T^2}{8}\,.
    \end{align}
\end{subequations}


\providecommand{\href}[2]{#2}\begingroup\raggedright\endgroup

\end{document}